\documentclass[letterpaper,twocolumn,10pt]{article}
\usepackage{usenix2019_v3}

\usepackage{tikz}
\usepackage{amsmath}

\usepackage[framemethod=TikZ]{mdframed}
\usepackage{amssymb}
\usepackage{amsthm}
\usepackage{graphicx}
\usepackage{algorithm}
\usepackage{algpseudocode}
\usepackage{stackengine}
\usepackage{scalerel}
\usepackage{amsmath}
\usepackage{centernot}
\usepackage{mathtools}
\usepackage{stmaryrd}

\usepackage{enumitem}
\usepackage{xcolor}
\usepackage{latexsym}
\usepackage{appendix}
\usepackage[colorinlistoftodos]{todonotes}
\graphicspath{{images/}}

\theoremstyle{plain}
\newtheorem{thm}{Theorem}
\theoremstyle{definition}
\newtheorem{lemma}{Lemma}
\newtheorem{defn}[thm]{Definition}

\newenvironment{myFrame}[1]
  {\mdfsetup{
    frametitle={\colorbox{white}{\space#1\space}},
    innertopmargin=10pt,
    frametitleaboveskip=-\ht\strutbox,
    frametitlealignment=\center,
    leftmargin = 10pt,
    rightmargin = 10pt,
    innertopmargin=10pt,
    innerbottommargin=10pt,
    innerrightmargin=10pt,
    innerleftmargin=10pt,
    }
  \begin{mdframed}%
  }
{\end{mdframed}}

\begin{document}

\date{}

\title{\Large \bf Cumulus: A BFT-based Sidechain Protocol for Off-chain Scaling}


\author{
{\rm Fangyu Gai} \\
School of Engineering \\
University of British Columbia \\
fangyu.gai@ubc.ca
\and
{\rm Jianyu Niu} \\
School of Engineering \\
University of British Columbia \\
jianyu.niu@ubc.ca
\and
{\rm Cesar Grajales} \\
School of Engineering \\
University of British Columbia \\
cesar.grajales@ubc.ca
\and
{\rm Mohammad Mussadiq Jalalzai} \\
School of Engineering \\
University of British Columbia \\
m.jalalzai@ubc.ca
\and
{\rm Chen Feng} \\
School of Engineering \\
University of British Columbia \\
chen.feng@ubc.ca
}

\maketitle

\begin{abstract}
Sidechain technology has been envisioned as a promising solution to accelerate today's public blockchains in terms of scalability and interoperability.
By relying on the mainchain for security, different sidechains can formulate their own rules to reach consensus.
Although the literature has considered the possibility of using consensus protocols in the sidechain, so far a tailor-made consensus protocol for sidechains with high performance and formal security proof has not been attempted.
To fill this gap, we introduce \textit{Cumulus}, a low overhead, highly efficient, security provable sidechain protocol.
Cumulus follows a BFT-based protocol and makes use of smart contracts to ensure that only one block proposed in the sidechain will be enforced on the mainchain in each round, thereby achieving consensus in an efficient manner.
We give a formal specification of Cumulus which ensures safety and liveness without any online requirements of clients.
For security analysis, we provide formal security definitions and proofs under \textit{Universally Composable Security} (UCS) model.
As proof of concept, we implement Cumulus and evaluate it in an Ethereum testnet.
\end{abstract}

\section{Introduction}

Recent years have witnessed a growing academic interest in the field of blockchain technology,
which is the cornerstone behind cryptocurrencies such as Bitcoin \cite{nakamoto2008bitcoin} and Ethereum \cite{wood2014ethereum}.
This technology has been envisioned as a trustless and decentralized platform on which developers are building various applications beyond cryptocurrencies.
However, today's public blockchains (often referred to as permissionless blockchains, such as Bitcoin and Ethereum) are hindered by the use of broadcast which significantly limits their scalability.
One promising approach to scaling blockchains is to enact transactions ``off-chain'', i.e. execute transactions off the blockchain, without compromising the security assumptions of the underlying blockchain.
Off-chain scaling significantly improves the scalability of the blockchain and also provides backward compatibility which is crucial for widely adopted blockchains whose consensus protocols are troublesome to change once deployed (might cause hard forks).

Off-chain solutions are also called layer-two protocols \cite{DBLP:journals/iacr/GudgeonMRMG19} which consider the existing blockchains as the base layer (layer-one).
Layer-two protocols typically assume two properties provided by the blockchain layer: (i) \textit{integrity}, meaning only valid transactions are added to the ledger, and (ii) \textit{eventual synchronicity}, which means a valid transaction will be eventually added to the ledger within a critical timeout.
Benefited by these, one off-chain proposal called payment channel networks (PCN) \cite{poon2016bitcoin} is proposed enabling two users to perform unlimited transactions off-chain only touching the mainchain during the deposit and withdrawal process.
While numerous contributions have been made to improve the performance of PCN \cite{miller2017sprites}\cite{perun2019}\cite{revive2017}, they still face multiple challenges such as costly routing, expensive channel setup and rebalance requirement.
Due to these obstacles, PCNs are mainly used in peer-to-peer micropayment scenarios. 

In this paper, we focus on another layer-two approach called \textit{sidechains} \cite{back2014enabling}, which has much wider application scenarios as different sidechains can offer different features.
Sidechains are originally proposed to solve interoperability issues via a two-way peg that allows users to transmit their assets between two stand-alone blockchains \cite{DBLP:journals/corr/DilleyPWPGF16}.
Another form of sidechains is constructed on the parent-child structure where the parent blockchain (referred to here as the mainchain) has multiple sidechains.
Thus, these sidechains are meant to depend on the mainchain to some extent \cite{pos-sidechain}, whereas they can run different consensus protocols (even without one) than the mainchain for scaling purpose.
We envision different sidechains are built on top of widely adopted blockchains supporting various blockchain-based applications and most of the transactions are performed within the sidechain except for a few inevitable on-chain operations (e.g., deposits, withdrawal, and periodically commitments), thereby off-chain scaling achieved.
We only consider this form of sidechains in the context.

\subsection{Motivation}

Plasma \cite{poon2017plasma} and NOCUST \cite{rami2018nocust} are two of the most promising sidechain approaches which employ the smart contract technology to achieve two-way peg.
They allow a centralized operator to manage the sidechain (or ``commit-chain'' in NOCUST) while guaranteeing that users always retain control over their assets.
On one hand, they can achieve the same magnitude of throughput as centralized fiat payment processing systems, such as MasterCard and Visa.
On the other hand, they ensure that users can always exit the sidechain if the operator is compromised as long as the mainchain is secure.
Despite the above two clear advantages, sidechain-based solutions are still in its infancy, facing several limitations.

\setitemize{leftmargin=*}
\begin{itemize}
    \item \textit{A failure-prone operator.} Although the operator can be trustless, its failure and misbehaviour still impose a critical impact on user experience.
    In particular, once the operator is compromised, so is the sidechain. Users have to perform mass-exit in order to keep their assets safe, which will probably cause congestion in the mainchain.
    \item \textit{Withholding attack.} This attack is originated from \textit{Selfish Mining}, and has emerged as a critical attack against PoW-based blockchains \cite{7728010} \cite{Eyal:2018:MEB:3234519.3212998} \cite{niu2019selfish}, where selfish miners withhold blocks for unfair mining competition.
    In the sidechain context, the withholding attack consists of the block proposer forging an arbitrary block, not propagating it to other nodes, but making the commitment to the mainchain.
    From the mainchain's perspective, the committed block is always accepted since it is too expensive to verify the correctness of its content.
    Current solutions include either instructing the users to exit the sidechain or forcing the proposer providing the data through the smart contract.
    However, both solutions require users to be online and the latter will involve multiple on-chain transactions.
    \item \textit{Online requirement.} Existing off-chain proposals, including sidechains and PCN, require the users to be online to receive the payment or to verify the block at the end of each block time for trustless operations.
    Otherwise, sidechain assets might be stolen.
    This requirement is sometimes impractical to fulfill.
    There are some potential solutions such as ``watch towers'' \cite{mccorry2018pisa} which enables users to delegate to a third party to perform some critical operations on their behalf.
    However, this introduces additional trust assumptions and costs.
\end{itemize}

\subsection{Our Solutions and Contributions}

In order to address the above issues, we propose to design new consensus algorithms for sidechains.
Although some existing work \cite{croman2016scaling}\cite{poon2017plasma}\cite{pos-sidechain} has discussed the possibility of consensus mechanisms for sidechain (e.g., PoS, PoA, PBFT, etc.), a consensus protocol specially designed for sidechains with formal security proof is still missing in the literature.
To fill this research gap, we present Cumulus: a low overhead, highly efficient, security provable sidechain protocol.
More specifically, Cumulus (i) allows organizations to run separate sidechains and provides various off-chain services to their users, (ii) provides \textit{instant finality} which means once a block is committed in the sidechain, transactions in that block will be finalized instantly, and (iii) gives a guarantee to users that their assets on sidechains are always safe without online requirements and they can always withdraw/exit with their coins back to the mainchain.
Our contributions can be summarized as follows:

\textbf{A BFT-based consensus algorithm specially designed for sidechain.}
The crux of designing a consensus algorithm for sidechains is how to make the mainchain aware of the sidechain updates safely and efficiently.
To solve this problem, we introduce a new algorithm called Smart-BFT which combines \textit{Synchronous Byzantine Agreement} (SBA) \cite{sba} and smart contract technologies to solve the well-known Byzantine consensus problem \cite{Lamport:1982:BGP:357172.357176} in the sidechain context.
Our main contribution in this section is to customize the SBA so that it can update the mainchain about the approval of each block.
Besides, it can detect any Byzantine behavior which may include submitting an invalid block or not submitting a block at all to the mainchain by the block proposer.
Thus, the Smart-BFT can provide clients with a trusted environment where they do not have to remain online to verify blocks.
The protocol tolerates $f$ Byzantine nodes among $n\geq2f+1$ parties to guarantee safety and liveness.

\textbf{Fair withdrawal/exit.}
Depositing money in the sidechain is obviously easy since users can easily issue a forward transfer (from the mainchain to the sidechain) to join the sidechain and every party can verify on-chain transactions.
However, it is difficult to keep fairness vice versa because of the opacity of off-chain transactions.
Fairness means that users (i) can only withdrawal/exit money within their balances, and (ii) during or after the withdrawal/exit process, they cannot spend the associated money.
In Cumulus, users withdraw/exit directly through the smart contract by providing appropriate proof of possessions.

\textbf{Security Model and Formal Proof.} To be able to rigorously analyze the security properties of our protocol, we employ a general-purpose definitional framework called the Universally Composable Security (UCS) framework by Canetti \cite{canetti2001universally}.
By using the UCS framework, we formally define the security properties of our protocol and provide detailed proof.

\subsection{Outline}

Section~\ref{sec:system-overview} describes the system at a high level.
Section~\ref{sec:sidechain} presents Cumulus in detail.
Section~\ref{sec:security} gives formal secure definition of Cumulus.
Section~\ref{sec:implementation} evaluates the performance and scalability.
Section~\ref{sec:related-work} discusses the related work and Section~\ref{sec:conclusion} concludes this paper.

\section{System Overview} \label{sec:system-overview}
In this section we first define key elements in our system.
We then present the system model including the communication, blockchain, as well as the adversaries and cryptographic primitives.
Finally, we introduce our system goals.

\subsection{Main Elements}

Cumulus involves the following four types of roles:

\textbf{Mainchain}: We assume a generic blockchain model as $MC$ where the blockchain can run arbitrary Turing-complete programs (i.e., smart contracts) \cite{DBLP:conf/sp/KosbaMSWP16}.
We also assume $MC$ is a trusted third party that ensures the correctness and availability of the data including smart contracts, accounts, balances, and transactions.

\textbf{Smart Contract}: A smart contract is a piece of code deployed on $\mathbb{MC}$ acting as the bridge between $\mathbb{MC}$ and the sidechain (denoted by $\mathbb{SC}$) and providing dispute-solving services.
We use $\mathbb{C}$ to denote a set of contracts designated to a specific $\mathbb{SC}$.
In order to meet specific requirements of different sidechains, each sidechain has a set of exclusive smart contracts containing a group of functions.
We assume that within $\Delta$ time, a transaction sent to $\mathbb{C}$ will be confirmed on $\mathbb{MC}$ which means within $\Delta$ time, the block containing the transaction will have sufficient confirmations by miners that the possibility for it to be replaced is negligible.

\textbf{Validators}: A set of validators $\mathcal{V}=\{V_1, V_2,...,V_n\}$ run validations and consensus algorithms in $\mathbb{SC}$, where $n$ is the number of validators.
Each of them is identified by a unique pair of public/private keys in the sidechain network.
Validators take turns to propose blocks, one block at an epoch.
During each epoch, the validator who is responsible to propose the block is called the \textit{leader}, while the rest are called \textit{followers}.
The order of rotation is prefixed.

\textbf{Clients}: A set of clients $\mathcal{C}=\{C_1, C_2,...,C_m\}$ are actual users of $\mathbb{SC}$, where $m$ is the number of clients inside $\mathbb{SC}$ network.
They can share the same public/private key pair between $\mathbb{MC}$ and $\mathbb{SC}$.
Clients issue transactions directly to validators.
In order to increase the success rate of acceptance, clients can multicast transactions to a group of validators they trust.

\subsection{System Model}

\subsubsection{Communication Model}

We assume our protocol under a synchronous communication network, where there is a known upper bound on the message transmission delay.
We assume all parties have access to $\mathbb{MC}$ and $\mathbb{C}$, which are always available.
All the transactions sent to $\mathbb{MC}$ and $\mathbb{C}$ will be finalized within time $\Delta$.
We assume parties can instantaneously acquire the status of $\mathbb{C}$ so that they are always aware of the current state of $SC$ (e.g., epoch number, the $id$ of the current leader).

\subsubsection{Adversary Model}

We consider the adversary can compromise all the clients and $f$ validators, where $f \leq \frac{n-1}{2}$.
Faulty parties may deviate from their normal behavior in arbitrary ways, e.g., hardware/software crash, nonsensical read and write to $\mathbb{MC}$, colluding with each other to cheat honest parties, etc.
However, we assume they do not have enough computation power and money to compromise $\mathbb{MC}$.
We also assume that all messages are signed by their senders and adversaries cannot break collision-resistant hashes and signatures used both in $\mathbb{MC}$ and $\mathbb{SC}$.

\subsubsection{Cryptographic Primitives}
Our protocol utilizes \textit{aggregate signatures} which enable distributed signing among $n$ participants $\{P_1,P_2,...,P_n\}$.
An aggregate signature scheme consists of four algorithms: \textsf{KeyGen}, \textsf{Sign}, \textsf{Combine}, and \textsf{Verify}.
The former two algorithms are the same as the standard signature scheme.
Algorithm \textsf{Combine} takes a vector of $n$ triples ($pk_i$, $m_i$, $\sigma_i$) as input, where $pk_i$ is the public-key for $C_i$, $m_i$ is the message, and $\sigma_i$ is the signature signed on $m_i$.
The output is a single aggregate signature, whose length is the same as a signature on a single message.
Algorithm \textsf{Verify} takes $n$ pairs ($pk_i$, $m_i$) and the aggregate signature as input and outputs ``True'' only if $\sigma$ was generated as an aggregate of $n$ valid signatures.

\subsection{System Goals} \label{sec:system-goals}

We emphasize that our scheme provides both \textit{safety} and \textit{liveness} assuming no more than $\lfloor\frac{n-1}{2}\rfloor$ validators are faulty without online requirements.
We now informally present the security and efficiency properties that Cumulus aims to provide, which are formally discussed in Sec.~\ref{sec:smartbft-safety} and Sec.~\ref{sec:ideal-function}.

\textit{Consensus on the sidechain.} Honest validators will commit the same block for each epoch (\textit{safety}).
In particular, an honest validator will only commit a block if it has a valid \textit{quorum certificate} ($QC$), which is an aggregate of signatures of $f+1$ validators.
Furthermore, a checkpoint submitted to $\mathbb{C}$ is only considered valid if it has a valid $QC$.
Otherwise, it will be challenged.
This means that every valid checkpoint finalized on $\mathbb{C}$ corresponds to a block that is committed in $\mathbb{SC}$ but not vice versa (\textit{weak consistency}).
A checkpoint is produced within $O(\tau+\Delta)$ time in optimistic cases and within $O(f(\tau+\Delta))$ in pessimistic cases, where $\tau$ is the time to commit a block on $\mathbb{SC}$ while $\Delta$ is that time for $\mathbb{MC}$ to confirm a submitted checkpoint.

\textit{Guaranteed fair withdrawal/exit.} The users of a sidechain are guaranteed that they can always withdraw a certain amount of money to the mainchain (within their balance), or exit the sidechain with all the assets from the latest \textit{settled} checkpoint (\textit{liveness}).
During or after the process, the users cannot spend the money they claimed.
A user can successfully withdrawal/exit within $O(\tau+\Delta)$ time in optimistic cases and $O(f(\tau+\Delta))$ in pessimistic cases, where $f$ is the number of Byzantine validators.

\section{The Cumulus Sidechain} \label{sec:sidechain}

In this section, we present the design details of Cumulus.
Similar to Ethereum \cite{wood2014ethereum}, Cumulus is also an account-based protocol, where each participant has an account representing their properties (e.g., balance, non-fungible tokens, etc.).
Therefore, the whole protocol can be viewed as a state machine.
For simplicity, in this paper, the state only refers to account balances.
The protocol begins with a genesis state and proceeds by executing transactions continuously until it terminates.

\subsection{Cumulus Overview}

\begin{figure}
\begin{center}
  \includegraphics[width=0.9\linewidth]{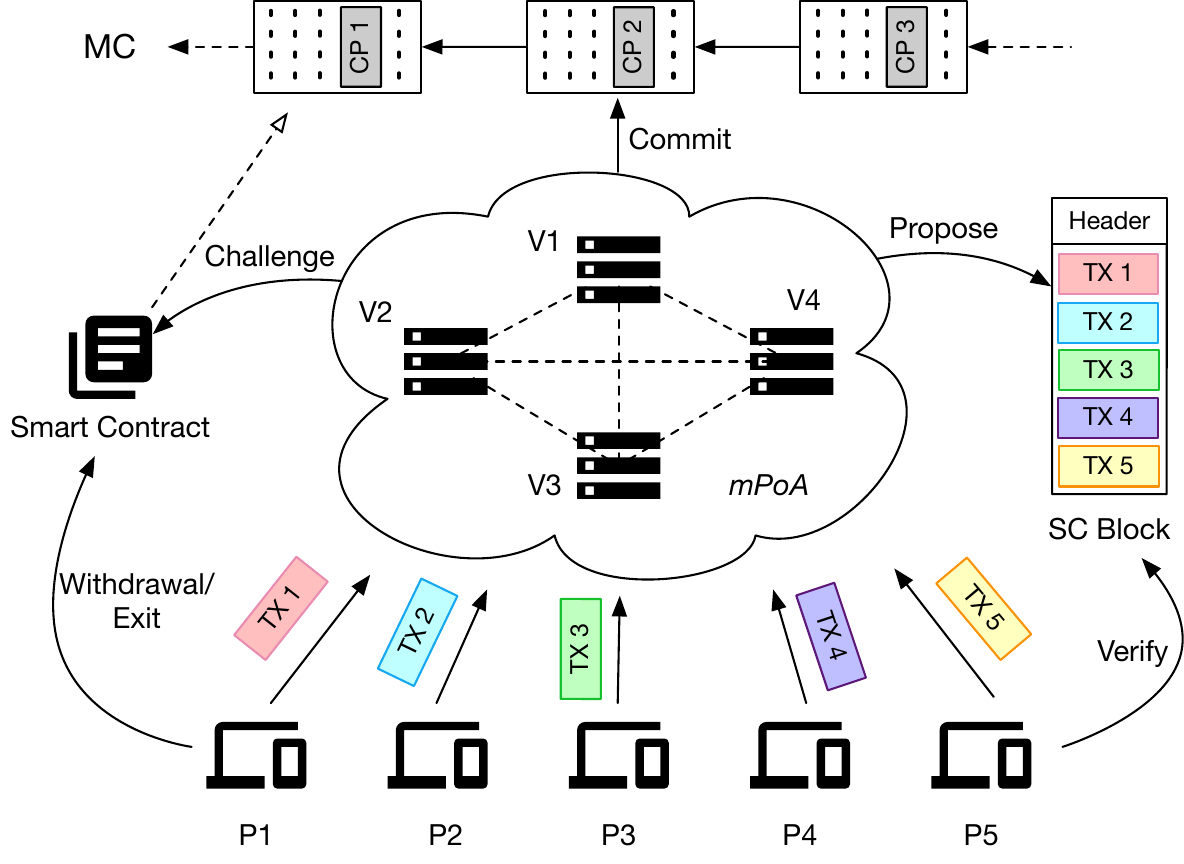}
  \caption{An overview of transaction flow in Cumulus.}
  \label{fig:overview}
\end{center}
\end{figure}

In Cumulus, we rely on a fixed set of validators $\mathcal{V}$ to maintain $\mathbb{SC}$.
The sidechain $\mathbb{SC}$ is set up by $\mathcal{V}$ deploying the smart contracts $\mathbb{C}$ that contain the public key as the identity of each validator for future verification.
The protocol proceeds in epochs.
Each epoch has a designated validator as the leader to collect signed transactions from clients, execute them, and package them along with the outcome into a block.
Then the leader proposes that block by broadcasting it to all the validators.
On receiving a block, a validator checks the validity of the block and sends a signed vote if the block is valid.
If the leader collects sufficient votes (more than $f$), then it aggregates the votes into a \textit{quorum certificate} ($QC$) that provides evident of receiving enough votes and commits the hash of the block along with the $QC$ to $\mathbb{C}$ as a \textit{checkpoint} (cf. Sec. ~\ref{sec:checkpoint}) to enforce the state of the epoch.
If a Byzantine leader submits an invalid checkpoint or does not submit any checkpoint (even if a valid $QC$ is reached), the leader can be \textit{challenged} and slashed if the challenge is successful (cf. ~\ref{sec:challenge}).
Clients can always withdraw/exit from the latest \textit{settled} checkpoint (cf. Sec. ~\ref{sec:withdrawl}).
Fig.~\ref{fig:overview} shows a system overview of Cumulus.
During this process, many Byzantine cases can happen.
We elaborate on our BFT protocol in the following section and discuss its safety and liveness in Sec. ~\ref{sec:smartbft-safety}.

\subsection{Smart-BFT} \label{sec:smart-bft}

In this section, we present the consensus engine of Cumulus, Smart-BFT, a synchronous BFT protocol tolerating $f$ faults among $n\geq 2f+1$.
Smart-BFT is built by combining smart contracts with the \textit{Synchronous Byzantine Agreement} (SBA) \cite{sba} without touching its consensus core.
This combination enables that: (i) once a block is committed, it will be updated on $MC$, (ii) leader election is achieved within smart contracts instead of an additional round of consensus, and (iii) Byzantine leaders can be easily detected and slashed.
The core idea of Smart-BFT is to guarantee all honest parties eventually agree on the same block which will be committed onto $MC$ by an honest leader for each epoch.
The algorithm for random leader election is beyond the scope of this paper.
Some potential solutions can be found in \cite{DBLP:journals/iacr/BonneauCG15, DBLP:journals/ccds/PierrotW18}.
For ease of exposition, we just assume that we follow a \textit{leader rotation} procedure which has been widely used to fairly distribute the responsibility of block production among validators \cite{soton411996}\cite{8123568}, and the order of validators taking the lead is predetermined in $\mathbb{C}$.
Each party can simply query $\mathbb{C}$ to learn the identity of the current leader.

Cumulus clients issue transactions to one or multiple validators who forward them to other validators through a gossip protocol.
Validators keep these transactions in their memory pool, and when a validator is selected by $\mathbb{C}$ as the leader $L_k$ for epoch $k$, the leader $L_k$ will choose a collection of transactions and execute them (update users' account according to the transactions).
Then $L_k$ packs the transactions along with the outcome into a block, denoted by $B_k$.
The protocol proceeds in epochs.
We now describe the protocol for epoch $k$ in detail.
Note that all messages are signed by the sender and broadcasting a message means that the sender sends the message to all parties including itself.
For simplicity, we omit to describe the verification of the signatures.

\begin{algorithm}[t!]
\caption{The Smart-BFT algorithm at $V_i$ for epoch $k$}\label{alg:smart-bft}
\begin{algorithmic}[1]
\State\textbf{Global State}:

\State\hspace{\algorithmicindent} $k$: the current epoch number
\State\hspace{\algorithmicindent} $L_k$: the public key of the leader for epoch $k$
\State\hspace{\algorithmicindent} $cp_{i}$: all the checkpoints for epoch from $0$ to $k-1$

\State \textbf{Local State}:
\State\hspace{\algorithmicindent} $T$: a set of transactions maintained in $V_i$'s memory
\State\hspace{\algorithmicindent} $B_{k-1}$: The block committed in epoch $k-1$
\State\hspace{\algorithmicindent} $QC_{B_{k-1}}$: The quorum certificate of $B_{k-1}$

\Statex \textcolor{blue}{$\triangleright$ \textsc{pre-prepare} phase}
\State as a \textbf{leader}:
\State\hspace{\algorithmicindent}
 $B_k\leftarrow$\textsf{execute($B_{k-1}, T$)}
\State\hspace{\algorithmicindent} broadcast (\underline{pre-prepare}, $B_k^L$)

\Statex \textcolor{blue}{$\triangleright$ \textsc{prepare} phase}
\State upon receipt of a valid \underline{pre-prepare} message:
\Statex\hspace{\algorithmicindent} broadcast (\underline{prepare}, $B_k$) \Comment{the hash of $B_k$}
\State wait for \underline{prepare} messages from other validators:
\State\hspace{\algorithmicindent} $\beta\leftarrow\{B_k^{\prime}|$ forwarded \underline{prepare} messages$\}$
\State\hspace{\algorithmicindent} \textbf{if} $\exists B_k^{\prime}\in\beta\neq B_k$ \textbf{then}
\State\hspace{\algorithmicindent}\hspace{\algorithmicindent} end this phase (equivocation observed)
\State\hspace{\algorithmicindent} \textbf{else if}
 $\forall B_k^{\prime}\in\beta= B_k$ and $|\beta|\geq f+1$ \textbf{then}
\State\hspace{\algorithmicindent}\hspace{\algorithmicindent} $QC_{B_k} =$ \textsf{aggregate}$(\beta)$

\Statex \textcolor{blue}{$\triangleright$ \textsc{commit} phase}
\State \textbf{if} $QC_{B_k}\neq \bot$
\State\hspace{\algorithmicindent} broadcast (\underline{lock}, $B_k$, $QC_{B_k}$) \Comment{the hash of $B_k$}
\State upon receipt of a valid message (\underline{lock}, $B_k$, $QC_{B_k}$):
\State\hspace{\algorithmicindent} \textbf{if} \textsf{verify}$(QC_{B_k})$ \textbf{then}
\State\hspace{\algorithmicindent}\hspace{\algorithmicindent} \textsf{commit$(B_k)$}
\State as a \textbf{leader}:
\State\hspace{\algorithmicindent} upon receipt of a set of valid \underline{lock} messages:
\State\hspace{\algorithmicindent}\hspace{\algorithmicindent} $\alpha\leftarrow$ $\{B_k |$ \textsf{matchMsg}$(B_k$, \underline{lock}$)\}$
\State\hspace{\algorithmicindent}\textbf{if} $|\alpha|\geq f+1$ \textbf{then}
\State\hspace{\algorithmicindent}\hspace{\algorithmicindent} $QC_{B_k} =$ \textsf{aggregate}$(\alpha)$

\Statex \textcolor{blue}{$\triangleright$ \textsc{submit} phase}
\State as a \textbf{leader}:
\State\hspace{\algorithmicindent} send (\underline{submit}, $B_k$, $QC_{B_k}$) to $\mathbb{C}$ \Comment{the hash of $B_k$}
\State wait for the \underline{submit} message to be confirmed in $\mathbb{C}$:
\State\hspace{\algorithmicindent} $k\leftarrow k+1$
\State\hspace{\algorithmicindent} $L_{k+1}, cp_{k}\leftarrow$ \textsf{query()}
\State\textbf{if} \textsf{verify($B_{k}$, $cp_{k}$) is \textsf{false}} \textbf{then}
\State\hspace{\algorithmicindent} send (\underline{challenge}, $L_k$) to $\mathbb{C}$

\end{algorithmic}
\end{algorithm}

The protocol given in Algorithm~\ref{alg:smart-bft} is described in separate phases of a single epoch.
The smart contracts $\mathbb{C}$ keep all the global state, such as the current epoch number, the identity of the current leader, and all of the committed checkpoints.
Every party can view the global state by simply querying $\mathbb{C}$ via \textsf{query()}.

\textbf{Phase 1 \textsc{pre-prepare}}. The leader $L_k$ starts epoch $k$ by executing local transactions based on the previous block $B_{k-1}$ that is committed in the previous epoch $k-1$ to build a new block $B_k$.
Then $L_k$ proposes $B_k$ by broadcasting it to all the followers via a \underline{pre-prepare} message.
Upon receipt of the proposed block, followers verify its validity (e.g., reject the block if the execution result of the block is incorrect).

\textbf{Phase 2 \textsc{prepare}}. All parties forward the received block (only the hash) to other parties by broadcasting a \underline{prepare} message.
Then they collect \underline{prepare} messages from other parties until the end of this phase.
Among all the blocks (only the hash) forwarded by others, if there exists one block $B_k^{\prime}$ and $B_k^{\prime}\neq B_k$, then $V_i$ will skip the rest of the phases of epoch $k$ because an equivocation is observed.
Otherwise, if $V_i$ receives $f+1$ \underline{prepare} messages all of which contain the same block $B_k^{\prime}$ and $B_k^{\prime}=B_k$, $V_i$ constructs $QC_{B_k}$ by aggregating all the signatures.

\textbf{Phase 3 \textsc{commit}}. If $QC_{B_k}\neq \bot$, $V_i$ locks the proposal of $B_k$ by broadcasting a \underline{commit} message to the rest of the parties.
At the end of this phase, if $V_i$ receives a valid \underline{commit} message in which the $QC_{B_k}$ is correct, it will commit $B_k$.
Meanwhile, the leader $L_k$ processes a set of \underline{lock} messages in all of which if $f+1$ blocks (only the hash) are the same as the one it proposed in the previous phase, the leader will aggregate the signatures into a new $QC_{B_k}^{\prime}$.

\textbf{Phase 4 \textsc{submit}}. The leader $L_k$ submits the block $B_k$ (only the hash) to $\mathbb{C}$ via a \underline{submit} message.
The smart contracts confirm the \underline{commit} message as $cp_k$ and enters the next epoch by changing the current leader.
At the same time, every party can verify the correctness of the checkpoint locally.
If $cp_k$ cannot match $B_k$ which is locked in the \textsc{commit} phase, at least one honest validator will challenge the invalid checkpoint by sending a \underline{challenge} message to $\mathbb{C}$.

\begin{figure}[h!]
\begin{center}
  \includegraphics[width=0.9\linewidth]{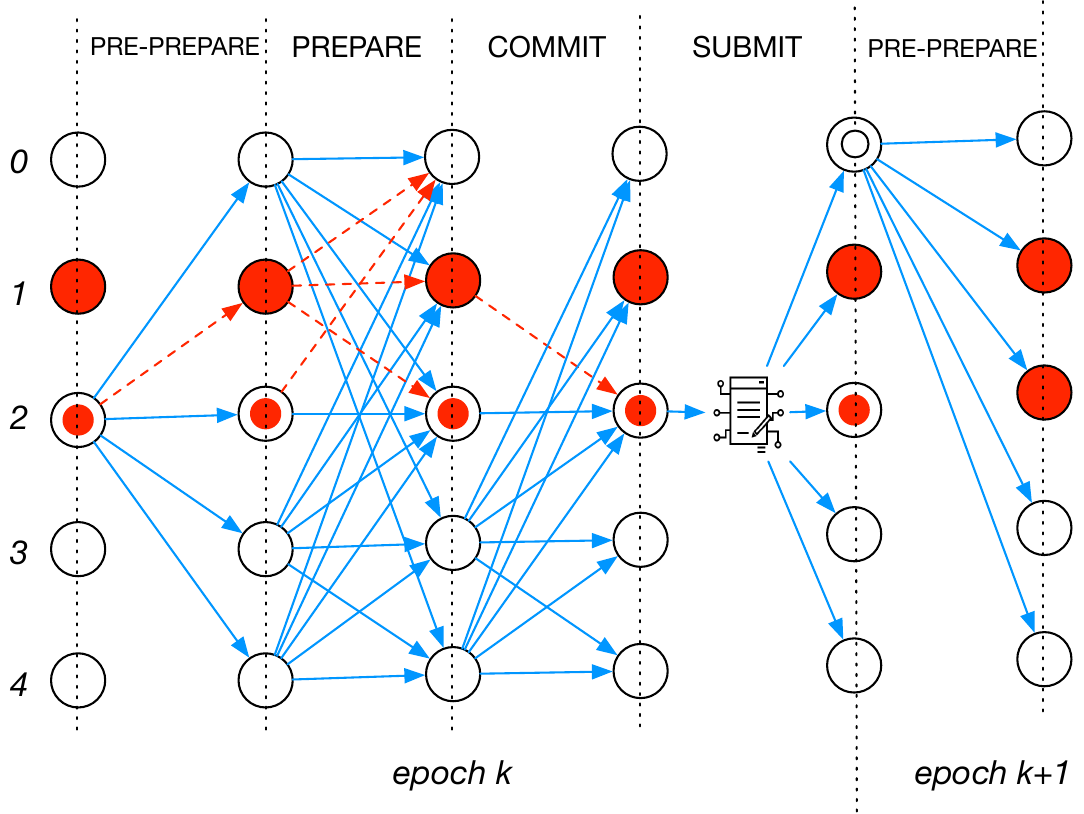}
  \caption{An example iteration of the Smart-BFT protocol where $f=2$ and $n=2f+1=5$.}
  \label{fig:smart-bft}
\end{center}
\end{figure}

Fig.~\ref{fig:smart-bft} shows an example iteration of how Smart-BFT protocol proceeds. In this example, we have $\mathcal{V}=\{V_0, V_1, V_2, V_3, V_4\}$, among which $V_1$ and $V_2$ are Byzantine and $V_2$ is the leader $L_k$ for epoch $k$.
1. (\textsc{Pre-prepare.}) The leader $L$ equivocates by sending one block (dashed red arrow) to $V_1$ and another different block (solid blue arrow) to other validators.
2. (\textsc{Prepare.}) Honest validators forward the received block (only the hash) to other validators while $V_1$ only multicasts the block to $V_0$ and $V_2$, and $V_2$ only sends the block to $V_0$.
Both of the honest parties $V_3$ and $V_4$ receive $3$ \underline{prepare} messages on the same proposal, which are used to construct a $QC$.
However, $V_0$ detects an equivocation and will not participate in the rest phases.
3. (\textsc{Commit}.) Honest parties $V_3$ and $V_4$ lock the block and broadcast \underline{Commit} messages to the rest of the parties while the Byzantine validator $V_1$ sends an arbitrary message (e.g., the red block) to $L_k$.
At the end of this phase, all honest validators $V_0$, $V_3$, and $V_4$ commit the blue block and the leader $L$ receives sufficient valid \underline{Commit} messages to generate a $QC$.
4. (\textsc{Submit}.) The leader submits the blue block (only the hash) along with some metadata to $\mathbb{C}$ and $\mathbb{C}$ confirms it, appends a new checkpoint, changes the leader to $V_0$, and enters the next epoch.

\subsection{\textit{Challenge} Mechanism} \label{sec:challenge}

We introduce a \textit{challenge} mechanism to deal with invalid or nil checkpoint submits so that Byzantine leaders can be slashed.
We present how it works by describing two Byzantine scenarios.
The first case is when the Byzantine leader crashes (or pretends to be off-line).
To keep liveness of Smart-BFT, a view-change should be triggered if no commitments arrive at $\mathbb{C}$ after a timeout.
We achieve this by setting a timer inside $\mathbb{C}$, which will be reset along with a commitment being confirmed in $\mathbb{C}$.
Once timeout, any parties can trigger the view-change (selecting a new leader) by simply sending a \underline{challenge} message to $\mathbb{C}$.

The second case is that Byzantine leaders may send arbitrary data to $\mathbb{C}$ as checkpoints no matter for collusive or motiveless reasons.
A checkpoint is considered to be faulty if it has invalid $QC$, which can be easily verified by any parties.
Faulty checkpoints devastate the security of a sidechain even though only the leader is Byzantine.
To address this issue, we need a ``challenge period'', which begins as soon as the checkpoint is confirmed and ends right after $\mathbb{C}$ confirms the next one.
In other words, the latest committed checkpoint is \textit{unsettled} until the next checkpoint is confirmed in $\mathbb{C}$.
We now describe how to challenge a faulty checkpoint.
We assume a faulty checkpoint $cp_k$ is newly confirmed in the epoch $k$ and the sidechain enters the next epoch $k+1$, at the beginning of which, honest validators will verify the $QC$ of the checkpoint.
Then any parties who find the incorrectness of the checkpoint can send a \underline{challenge} message to $\mathbb{C}$ to trigger the verification within $\mathbb{C}$.
Then $\mathbb{C}$ runs the verification, and the incorrect checkpoint will be revoked and the compromised leader will be slashed.
Note that the challenge process of the checkpoint $k$ and the consensus process for the epoch $k+1$ are concurrent.
Note that in this paper, we only consider leader slashing.
Adding new validators to keep the committee size above a safe threshold is also important.
We leave this for future work.

\subsection{Checkpoint} \label{sec:checkpoint}

Checkpoints are used for withdrawal/exit elaborated in Sec.~\ref{sec:withdrawl}. Each checkpoint consists of a hash of the committed block header, the signature of the proposer, an \textit{aggregate signature}, and a \textit{bit vector}.
Aggregate signatures are used as evidence that the committed block is approved by sufficient validators.
There are several signature constructions to achieve signature aggregation \cite{lysyanskaya}\cite{boldyreva}\cite{boneh}.
To reduce the length of a checkpoint, we apply the one based on the short signature scheme of Boneh, Lynn, and Shacham \cite{shortsignature}.
The \textit{bit vector} represents whether or not the indexed validator has signed the \textit{approval}.
For example, we have $5$ validators in total, indexed by $V_0,V_1,V_2,V_3,V_4$.
The \textit{bit vector} of a certain checkpoint is $11$ whose binary format is $01011$, meaning that $v_1,v_3,v_4$ signed the \textit{approvals}.
Let $cp_i=\{H(B_i),QC(H(B_i)),\text{\textsc{index}}\}$ denote the $i$th checkpoint, where $H(B_i)$ is the hash of the header of block $B_i$, $QC(H(B_i))$ is a constant-sized aggregate signature on $H(B_i)$, and \textsc{index} is an integer which contains the indexes of a list of validators that have signed the $QC(H(B_i))$.

Checkpoints have three different status: \textit{unsettled}, \textit{revoked}, and \textit{settled}.
The status of the latest checkpoint $cp_k$ (assume the current epoch is $k$) is \textit{unsettled} since it is likely to be \textit{revoked} if challenged.
If it is successfully challenged in epoch $k+1$, the smart contract $\mathbb{C}$ will revoke the checkpoint by setting $cp_k=\bot$.
Otherwise, when the next checkpoint $cp_{k+1}$ comes, its status will be set to \textit{settled} because no \underline{challenge} messages are received by $\mathbb{C}$ during a challenge period, which means clients are safe to withdraw/exit through $cp_k$.
Meanwhile, the newly confirmed checkpoint $cp_{k+1}$ will be set to \textit{unsettled}.

\subsection{Smart-BFT's Safety and Liveness} \label{sec:smartbft-safety}
We now present the underlying safety and liveness properties of Smart-BFT based on Theorem 2 and Theorem 3 from SBA \cite{sba} since its consensus core is not touched.

\noindent \textbf{Safety.}
The safety property provided by the algorithm is twofold.
First, the Smart-BFT ensures agreement, which means that all honest validators commit on the same value.
Second, the Smart-BFT provides weak consistency with the smart contract, which means that every settled block (checkpoint) on $\mathbb{C}$ is committed by all honest parties.

\begin{thm} \label{theorem:1}
(Agreement inherited from Theorem $2$ in \cite{sba}).
If two honest validators commit on $B_k$ and $B_k^{\prime}$ in epoch $k$ respectively, then $B_k^{\prime} = B_k$.
\end{thm}

\begin{lemma} \label{lemma:1}
If a checkpoint $cp_{k}$ containing block $B_k$ is settled on $\mathbb{C}$ and an honest party committed block $B_k^{\prime}$, then $B_k=B_k^{\prime}$.
\end{lemma} 

\begin{proof} 
Suppose for contradiction that $B_k \neq B_k^{\prime}$.
The checkpoint $cp_k$ being settled on $\mathbb{C}$  means that the leader has received $f+1$ \underline{commit} messages for $B_k^{\prime}$ because otherwise it will be challenged by at least an honest validator.
Due to Theorem \ref{theorem:1}, all of the honest parties will commit the block $B_k$, which means that no honest parties will send \underline{commit} messages for block $B_k^{\prime}$.
Consequently, it is impossible for the leader to receive $f+1$ valid signatures for $B_k^{\prime}$.
Thus, once $B_k^{\prime}$ is submitted to $\mathbb{C}$ at epoch $k$, it will be challenged by at least one honest party, and then be revoked by the end of epoch $k+1$, a contradiction.
\end{proof}

\begin{lemma} \label{lemma:2}
If an honest validator commits on $B_k$ and the leader submits $B_k^{\prime}$ onto $\mathbb{C}$ ($B_k\neq B_k^{\prime}$) as a checkpoint $cp_k^{\prime}$ at epoch $k$, then $cp_k$ will be revoked at the end of epoch $k+1$.
\end{lemma}

\begin{proof}
Due to Theorem \ref{theorem:1}, it is impossible for the leader to gain $f+1$ valid signatures to construct a $QC$ for block $B_k^{\prime}$, hence after the checkpoint $cp_k^{\prime}$ is confirmed by $\mathbb{C}$, at least one honest validator will challenge $cp_k^{\prime}$, which will be revoked at the end of epoch $k+1$.
\end{proof}

\begin{thm} \label{theorem:2}
(Weak consistency). If a block $B_k$ is committed by all honest validators at epoch $k$, then the checkpoint $cp_k^{\prime}$ will be either revoked or settled with $B_k=B_k^{\prime}$ at epoch $k+1$.
\end{thm} 

\begin{proof}
The proof is straightforward from Lemma \ref{lemma:1} and Lemma \ref{lemma:2}.
If the submitted block $B_k^{\prime}=B_k$, it will be settled on $\mathbb{C}$ at the end of epoch $k+1$ (by Lemma \ref{lemma:1}).
Otherwise, it will be revoked at the end of epoch $k+1$ (by Lemma \ref{lemma:2}).
\end{proof}

\noindent \textbf{Liveness.} The Smart-BFT guarantees two-level liveness.
The first level (termination) is that if the leader is honest, all honest validators are guaranteed to commit by the end of that epoch.
The second level is that a settled checkpoint eventually appears on $\mathbb{C}$ in finite time so that clients can withdraw/exit.

\begin{thm} \label{theorem:3}
(Termination inherited from Theorem $3$ in \cite{sba}.)
If the leader $L_k$ in epoch $k$ is honest, then every honest party commits by the end of epoch k.
\end{thm}

\begin{lemma} \label{lemma:3}
If an honest leader $L_k$ in epoch $k$ submits a block $B_k$ onto $\mathbb{C}$ as a checkpoint $cp_k$, then it will be settled by the end of epoch $k+1$.
\end{lemma} 

\begin{proof}
By Theorem \ref{theorem:3}, all honest validators commit on block $B_k$ which means the $QC$ of the checkpoint $cp_k$ is aggregated from $f+1$ valid signatures.
Therefore, no parties can successfully challenge $cp_k$, and hence, it will be settled when the next checkpoint comes.
\end{proof}

\begin{lemma} \label{lemma:4}
If the leader is being Byzantine, then a view-change will be triggered if at least one honest observes the Byzantine behavior.
\end{lemma}

\begin{proof}
At a high level, a Byzantine leader can perform Byzantine behavior by (i) not submitting anything to $\mathbb{C}$ or (ii) submitting an invalid checkpoint.
For the first case, every time a view-change happens, the smart contract $\mathbb{C}$ will reset a timer.
If $\mathbb{C}$ has not received anything after timeouts, at least one honest validator will trigger the view-change.
For the second case, at least one honest validator will challenge the checkpoint as soon as it is confirmed in $\mathbb{C}$, and this checkpoint will be revoked after the verification performed in $\mathbb{C}$.
\end{proof}

\begin{thm} \label{theorem:4}
(Liveness) A settled checkpoint emerges within $f+1$ epochs.
\end{thm}

\begin{proof}
First, we consider an honest-leader case.
If the leader is honest, due to Lemma \ref{lemma:3}, the submitted checkpoint will be settled by the end of the next epoch.
Otherwise, due to Lemma \ref{lemma:4}, either the leader will be changed by a view-change or the submitted checkpoint will be revoked by $\mathbb{C}$ by the end of the next epoch.
In the worst case, the smart contract $\mathbb{C}$ selects $f$ consecutive Byzantine leaders.
This means an honest leader will be selected after at most $f$ epochs (Byzantine leaders will be slashed) and by Lemma \ref{lemma:3} again, a valid checkpoint will be settled after at most $f+1$ epochs.
\end{proof}

\subsection{Block Construction} \label{sec:block}

A block in our protocol contains all the transactions issued in $\mathbb{SC}$ over a specific epoch\footnote{Transactions may be assigned to the next epoch due to network latency and Byzantine validators may arbitrarily discard transactions.}, as well as the updated states (i.e., balances) after executing those transactions.
As such, a block is a basic unit of a state transition in $\mathbb{SC}$.
More precisely, after a block is published, either all of the transactions completed (all transactions are computed correctly), or they fail, and the state of $\mathbb{SC}$ remains unchanged.

\begin{figure}[h!]
\begin{center}
  \includegraphics[width=1.0\linewidth]{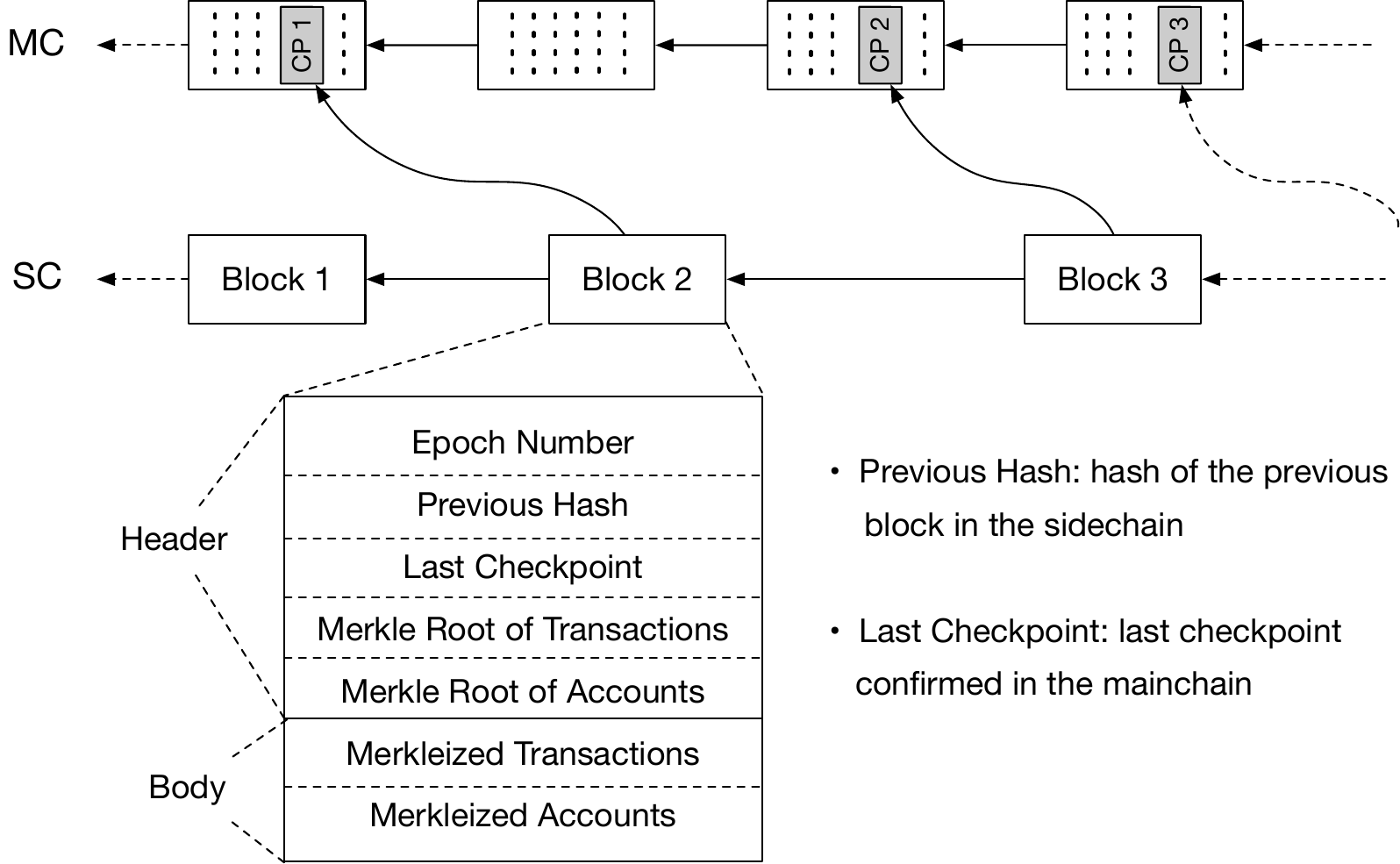}
  \caption{The construction of $\mathbb{SC}$ blocks.}
  \label{fig:block}
\end{center}
\end{figure}

The design goal of constructing a block in $\mathbb{SC}$ is
(i) to include all the information needed for a participant to withdraw or exit,
(ii) that when a participant wants to withdraw or exit, $\mathbb{C}$ should be able to efficiently verify the proof of possession of the funds,
and (iii) that all of the finalized blocks should follow a strict order according to when they are confirmed in $\mathbb{MC}$.
To fulfill those requirements, the blocks consist of a list of transactions, a list of updated accounts, along with the metadata required for verification.
Fig.~\ref{fig:block} depicts how $\mathbb{SC}$ blocks are constructed.

\textbf{Block header}. The block header contains metadata including the epoch number, the hash of the previous block, and the merkle roots of transactions and balances of accounts.

\textbf{Block Body}. The block body consists of two merkle trees, a transaction tree, and an account tree.
The transaction tree is a regular \textit{Merkle Tree} (MT) used in Bitcoin \cite{nakamoto2008bitcoin}, where hashes of transactions form leaf nodes and every non-leaf node is labeled with the hash of its two child nodes.
Verifiers can efficiently and securely verify the existence of a leaf node in a large data structure by providing an MT proof which is the path consisting of all the hashes going up from the leaf node to the root.
For the account tree, we introduce the \textit{Merkle Patricia tree} (MPT) structure -- used in Ethereum \cite{wood2014ethereum}, to merkleize the accounts.
In the context, the account id is the key and the balance of the account is the value in the MPT.
Verifiers can efficiently and securely verify the existence of a key and the correctness of the expected value to the key using MPT proof.

\subsection{Batching Blocks} \label{sec:batching}

In Smart-BFT, the time for each epoch is $\Gamma=\tau+\Delta$, where $\tau$ is the time for honest validators to commit a block while $\Delta$ is the blockchain response time and since $\Delta\gg\tau$, $\Gamma$ hinges on $\Delta$.
To avoid being stalled while waiting for the response from $\mathbb{MC}$, Smart-BFT can process blocks in batches while waiting and only submit the hash of the last block as the checkpoint to $\mathbb{C}$.
For example if the batch size $\mathcal{Z} = 10$, then the leader must submit the hash of blocks with the height $h\in\{h | h \mod 10=0\}$ (i.e., $0,10,20,30,...$).
Since each block points to the previous block using its hash, altering a single bit in any block within the batch will alter the hash of the final block.
Therefore, submitting the hash of the final block will suffice to prove the validity of previous blocks within the batch.

\subsection{Withdrawal/Exit} \label{sec:withdrawl}

The withdrawal/exit\footnote{The difference between a withdrawal and an exit is that a participant withdraws a certain amount of money equal or less than the balance without notifying any malicious behaviors (from a participant's view), while a participant exit with all of his money left in $\mathbb{SC}$ when some obvious maliciousness happens.} protocol can be viewed as a cross-chain protocol to transfer assets from $\mathbb{SC}$ to $\mathbb{MC}$.
This part is more complex than deposit (i.e., transferring assets reversely) because the updates in $\mathbb{SC}$ are unknown to $\mathbb{MC}$.
Therefore, participants have to provide sufficient proof to $\mathbb{MC}$ about the funds they want to withdraw.
Cumulus requires clients to withdraw or exit by sending a withdrawal/exit request directly to $\mathbb{C}$.
The rationale is that it is impractical for a leader to submit so many on-chain transactions to the $\mathbb{C}$ at a time, which probably will cause congestion in $MC$.
The withdrawal/exit request contains the value to withdraw and a \textit{Proof of Possession} (PoP) which is to prove a participant owns a certain balance of assets in $\mathbb{SC}$.
A standard PoP consists of the following components:
\begin{itemize}
    \item \textit{Block Header.} It is a regular block header that should be the same with the block header in the latest \textit{settled} checkpoint.
    \item \textit{Balance.} The latest balance of the account of the client.
    \item \textit{Merkle Path.} It is for $\mathbb{C}$ to verify whether the provided balance matches the expected value in the account tree with the MPT root included in the block header field.
\end{itemize}

Another concern in designing a secure withdrawal/exit protocol is that when a client submits a withdrawal/exit request to $\mathbb{C}$, the underlying PoW blockchain (i.e., Ethereum) needs some time to confirm the transaction (with high probability).
During this period, malicious users might spend the money that has been submitted for withdrawal/exit, thereby achieving double spending.
To avoid this issue, we propose a locking mechanism called ``claim-and-redeem''.
In a nutshell, users can only withdraw/exit from the penultimate checkpoint and the requester must ``claim'' it at least two epochs before the ``redeem'' point.
The rationale behind this is that (i) the users can only withdraw/exit from the latest settled checkpoint, which is the penultimate checkpoint, because the latest checkpoint can be repealed because of challenge and before the real withdrawal/exit, and (ii) after the ``claim'', honest validators will lock the account of the client in $\mathbb{C}$ until the client ``redeems'' the funds some point in the future.

\begin{figure}[h!]
\begin{center}
  \includegraphics[width=1.0\linewidth]{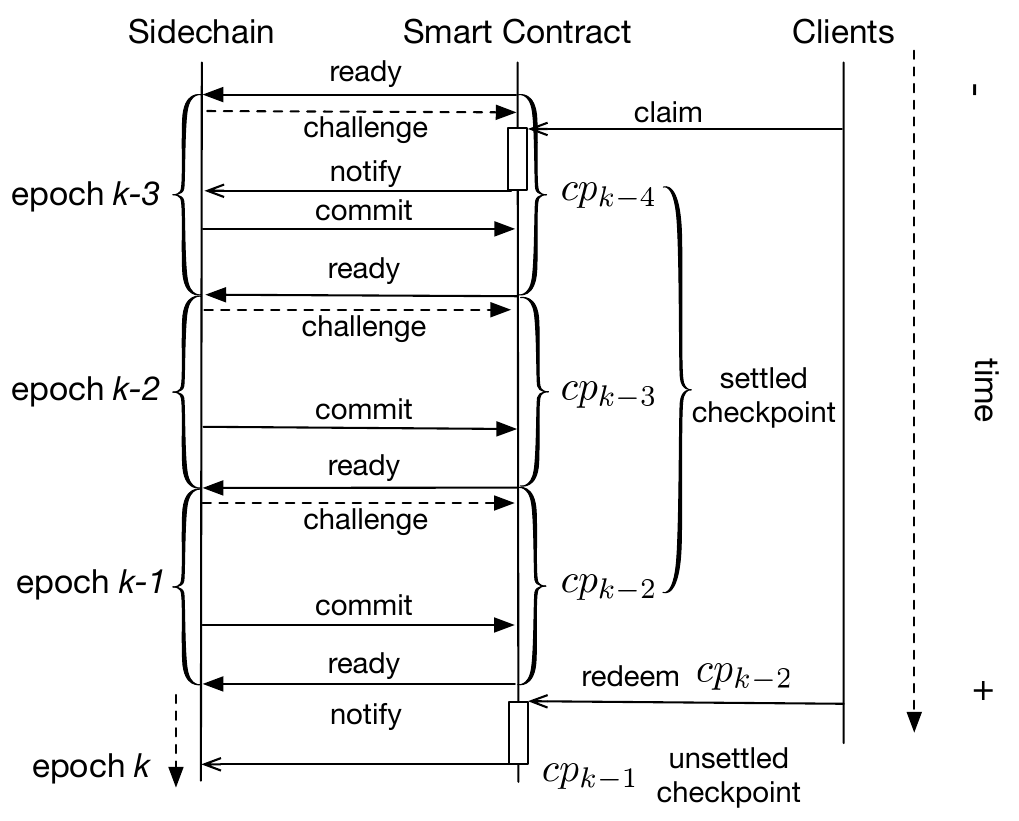}
  \caption{Claim-and-redeem mechanism. We assume that the ``claim'' transaction is sent at the beginning of epoch $k-3$ and confirmed in $\mathbb{C}$ by the end of the same epoch. In real life, the ``claim'' message may be sent several epochs earlier than it is confirmed in $\mathbb{C}$.}
  \label{fig:withdraw-exit}
\end{center}
\end{figure}

Fig.~\ref{fig:withdraw-exit} depicts how claim-and-redeem works.
For example, we have $3$ settled checkpoints $cp_{k-4}$, $cp_{k-3}$, and $cp_{k-2}$ committed during epoch $k-4$, epoch $k-3$, and epoch $k-2$, respectively.
Epoch $k$ is the current epoch, therefore $cp_{k-1}$ is unsettled.
A client issued a \textit{claim} transaction to $\mathbb{C}$ during epoch $k-3$ given that epoch $k$ is the current epoch.
Honest validators in $\mathbb{SC}$ acknowledged this request during the same epoch and would not accept any transactions in $\mathbb{SC}$ unless they saw a ``redeem'' transaction confirmed in $\mathbb{C}$ some point in the future.
Then, the client must wait for at least 2 epochs to redeem the withdrawal/exit.
Now we explain why at least 2 epochs waiting is necessary.
When honest validators first saw the claim transaction confirmed in $\mathbb{C}$, they will lock the account (rejecting any transactions related to the claimer) of that client immediately, but they might have processed some transactions related to the claimer.
Only from the beginning of epoch $k-2$, honest validators can guarantee that no transactions from the claimer will be processed until the redeeming point.
By the protocol proceeds to epoch $k$, $cp_{k-2}$ is the highest settled checkpoint and there have been two epochs after the claim.
Therefore, $\mathbb{C}$ will approve the redeem request on checkpoint $cp_{k-2}$ submitted during the current epoch $k$, after which validators will unlock the client's account.

\section{Formal Security Definition of Cumulus}
\label{sec:security}

\begin{figure*}[h!]

\begin{myFrame}{\textbf{Ideal Functionality $\mathcal{F}$}}

\begin{center}
	(I) \underline{\textbf{Deposit}}
\end{center}

A. Upon receiving $(\texttt{depositRequest}, \text{id}_{C_i^*},x)$ from $\mathcal{E}$, within $\Delta$ time remove $x$ coins from $C_i$'s account on the ledger $\mathcal{L}(\Delta)$ and:

\begin{enumerate}
    \item If $\text{id}_{C_m^*} \notin \mathcal{C}^*$, invoke a dummy machine $C_m^*$ for the client, update $\mathcal{C}^*:= \mathcal{C}^*\cup \{\text{id}_{C_m^*}\}$, update the balance for the client $B(\text{id}_{C_i^*}):=x$.
    Else, if $\text{id}_{C_i} \in \mathcal{C}$, send $(\texttt{existingClient})$ to $\mathcal{E}$. 
\end{enumerate} 

B. Upon receiving $(\texttt{turnToTalk})$ for a deposit of an amount $x$ through $\mathcal{C}_i^*$:

\begin{enumerate} 
    \item Update the balance for the client $B(\text{id}_{C_i^*}):=B(\text{id}_{C_i^*})+x$ and send $(\texttt{endOfTurn})$ to $\mathcal{E}$ through $C_i^*$.
\end{enumerate}

\begin{center}
	(II) \underline{\textbf{Transaction}}
\end{center}

A. Upon receiving $(\texttt{turnToTransfer})$ from $\mathcal{E}$ through $C_i$ to transfer an amount $x$ to $C_j^*$:

\begin{enumerate}
    \item Verify if $B_l(\text{id}_{C_i^*},e) \geq x$.
    If so, update the balance for the client $B(\text{id}_{C_i^*}):=B(\text{id}_{C_i^*})-x$ and  $B(\text{id}_{C_j^*}):=B(\text{id}_{C_j^*})+x$, the transactions queue $Q_{txn}:=Q_{txn} \cup \{txn\}$.
    Repeat this $n$ times to store an internal registry for each validator.
    \item If $t_{txn} \geq \tau$, randomly select the $j$th registry and compare it against the rest.
    If more than $f+1$ registries agree with the $j$th (including $j$th itself), store the content of $j$th registry as a block $B_k$.
    Otherwise, select another registry until the condition is met.
    After that, send $(\texttt{blockCommitted})$ to $\mathcal{E}$ within $O(\tau)$ time in optimistic cases and within ($O(f(\tau+\Delta))$ time in pessimistic cases, where leaders fail in succession).
    Then set the checkpoint $cp_k.status=unsettled$ and $cp_{k-1}.status=settled$ in $\mathbb{C}$ within $\Delta$ time.
\end{enumerate}

\begin{center}
	(III) \underline{\textbf{Withdraw/Exit}}
\end{center}

A. Upon receiving $(\texttt{turnToTalk})$ for a claim from $\mathcal{E}$ through $C_i$:
\begin{enumerate}
    \item Freeze $C_i$'s sidechain account, set a timer $\Delta_{redeem}^i=3\Gamma$, and send $(\texttt{accountFozen})$ to $\mathcal{E}$.
\end{enumerate}

B. Upon receiving $(\texttt{turnToTalk})$ for a redeem of an amount $x$ from $\mathcal{E}$ through $C_i$:

\begin{enumerate} 
    \item If $\Delta_{redeem}^i$ has expired, verify if $B(\text{id}_{C_i^*}) \geq x$. If so, transfer $x$ coins to $C_i^*$ account on ledger $\mathcal{L}$ after $\Delta$ time, update the balance for the client $B(\text{id}_{C_i^*}):=B(\text{id}_{C_i^*})-x$.
    Send $(\texttt{fundsRedeemed})$ to $\mathcal{E}$.
\end{enumerate}

\begin{center}
	(IV) \underline{\textbf{Challenge}}
\end{center}

A. Upon receiving $(\texttt{blockCommitted})$, $\mathcal{E}$ starts giving sequentially the turn to the validators and clients to request operations. Upon receiving $(\texttt{turnToTalk})$ from $\mathcal{E}$ through $V_k^*$ for a challenge for an invalid checkpoint:

\begin{enumerate}
    \item If the challenged checkpoint $cp_k.status=unsettled$, run the verification of $cp_k$.
    If $cp_k$ is valid, then set $cp_k.status:=settled$.
    Otherwise, update $\mathcal{V}:=\mathcal{V} \setminus \{V_l^*\}$ (the machine that is the leader of the epoch), set $cp_k:=\bot$, and send $(\texttt{checkpointRevoked})$ to $\mathcal{E}$.
\end{enumerate}

\begin{center}
	(V) \underline{\textbf{Restart}}
\end{center}

A. Upon receiving ANY message and the timer $\Gamma$ has expired:

\begin{enumerate}
    \item Reset the timer. Set $cp_k.status=settled$, increase the epoch $e:=e+1$ in $\mathbb{C}$, and send $(\texttt{ready})$ to $\mathcal{E}$ to start a new epoch.
\end{enumerate}

\end{myFrame}

\caption{Ideal functionality $\mathcal{F}^{\mathcal{L}(\Delta)}_{sc}(\mathcal{V},\mathbb{C})$ for the Sidechain in Cumulus.}
\label{fig:ideal-functionality}
\end{figure*}

\subsection{UC Security Model}

In this section, we formally present our sidechain using the Universally Composable Security (UCS) framework. In particular, we follow the work of \cite{general} applying a synchronous version of the global UC framework (GUC) \cite{guc}, an extension of the standard UC model allowing for a globally available set-up.
Under this model, all the deposits (from $\mathbb{MC}$ to $\mathbb{SC}$) and the withdrawal/exit (from $\mathbb{SC}$ to $\mathbb{MC}$) are handled via a global ideal functionality $\mathcal{L}(\Delta)$, the state of which is globally accessible by all parties.
$\mathcal{L}(\Delta)$  can freely add and remove money in user's accounts and the parameter $\Delta$ models that any interaction with $\mathcal{L}(\Delta)$ has a maximal delay $\Delta$.
A full definition of $\mathcal{L}(\Delta)$ can be found in \cite{dziembowski2018foundations}.

For a protocol to fulfill its intended security goals, two different protocols have to be designed: one representing an idealized model and one representing a real implementation. The ideal protocol will encompass all the functionality as a single entity, operating as desired, either normally or under any kind of attack. In the real protocol, each element is modeled as a separate computing entity. The security goal is achieved if any environment in which the protocol is deployed, cannot distinguish if it has interacted with the ideal model, or the realistic implementation.

We denote the ideal protocol with $\mathcal{I}$, and the real protocol with $\Pi$. The random variable that represents the output of the protocol $\Pi$ in conjunction with an adversary $\mathcal{A}$ and an environment $\mathcal{E}$, with inputs $k$ and $z$, is denoted by $\text{EXEC}_{{\Pi}, \mathcal{A}, \mathcal{E}}(k,z)$. To avoid complicating the notation, $\text{EXEC}_{{\Pi}, \mathcal{A}, \mathcal{E}}$ will represent the ensemble of probability distributions, for all possible choices of $k$ and $z$. With this, we can state our security definition.

\begin{defn} \label{EmulationDef}
The concept of security of the protocol $\Pi$ is achieved if for any adversary $\mathcal{A}$ and any environment $\mathcal{E}$, we can always find an adversary $\mathcal{S}$ (typically referred to as the ``simulator'') such that:

\begin{equation}
\text{EXEC}_{{\Pi}, \mathcal{A}, \mathcal{E}} \approx \text{EXEC}_{{\mathcal{I}}, \mathcal{S}, \mathcal{E}}.
\end{equation} 

\end{defn}

The ensembles of the execution of the protocol with environment $\mathcal{E}$ and adversary $\mathcal{A}$ are indistinguishable from the ensembles of the execution of the ideal functionality $\mathcal{I}$ with the same environment $\mathcal{E}$ and adversary $\mathcal{S}$. Part of the security proof is to build a simulator $\mathcal{S}$ for any given adversary $\mathcal{A}$. Whenever the conditions of Definition \ref{EmulationDef} are met, we say that $\Pi$ UC-realizes the ideal $\mathcal{I}$.

\begin{thm} \label{SecurityThm}
There exists a protocol $\Pi$, such that it UC-realizes the ideal protocol $\mathcal{I}$ for the ideal functionality $\mathcal{F}$.
\end{thm}


\subsection{The Ideal Functionality} \label{sec:ideal-function}

This section presents the ideal functionality for the protocol.
The UCS framework requires that, for the ideal protocol, the functionality of the whole system to be performed by a single entity.
To keep the protocol transparent to any environments or adversaries, the system must still have computing entities representing all the elements that are part of the protocol, but just as ``dumb machines'' that forward incoming messages to the functionality to perform what was requested, and forward the output to the environment $\mathcal{E}$.

Let $\mathcal{F}^{\mathcal{L}(\Delta)}_{sc}(\mathcal{V},\mathbb{C})$ denote the ideal functionality of $\mathbb{SC}$, where $\mathcal{V}$ is a set with a prefixed number of validators, $\mathbb{C}$ is a set of contracts deployed on  $\mathbb{MC}$ and $\mathcal{L}(\Delta)$ is the underlying ledger with delay $\Delta$.
The ideal functionality encompass all the functions of $V_1, \ldots, V_n$, $C_1, \ldots, C_m$, $\mathbb{C}$ and $\mathcal{L}$. It communicates with $\mathcal{E}$ via dummy parties $V_1^*, \ldots, V_n^*$, $C_1^*, \ldots, C_m^*$. For a fixed time $\Gamma=\tau+\Delta$, $\mathcal{E}$ gives turns to the clients to make deposits (or join), make withdraws, exit or challenge the checkpoint. It also gives turns to the validators to also challenge the checkpoint. Assume the epoch number is $k$.
To keep the model as simple as possible, we exclude transaction fees in our modeling.
The sidechain functionalities are specified in Fig.~\ref{fig:ideal-functionality}.
The \textbf{Deposit} functionality allows clients to join $\mathbb{SC}$ and increase their balance.
The \textbf{Transaction} functionality receives transactions from clients and instructs validators to commit blocks and submit checkpoints.
The \textbf{Withdraw/Exit} functionality provides two interfaces for claim, and redemption respectively.
The redemption request would only be processed if the claim request was confirmed at least $3$ epochs before the current epoch.
The \textbf{Challenge} functionality allows validators to question the unsettled checkpoint.
If the checkpoint is invalid, the relevant leader will be slashed.
The ideal functionality $\mathcal{F}^{\mathcal{L}(\Delta)}_{sc}(\mathcal{V},\mathbb{C})$ will enter the next epoch when it receives any messages after the timer $\Gamma$ expires.

We now discuss how the ideal functionality $\mathcal{F}^{\mathcal{L}(\Delta)}_{sc}(\mathcal{V},\mathbb{C})$ ensures the security and efficiency properties from Sec.~\ref{sec:system-goals}.

\textit{Consensus on the sidechain.}
The \textbf{Deposit} functionality guarantees the consensus on the genesis state of all the clients.
The \textit{safety} is achieved by the \textbf{Transaction} functionality sending the $(\texttt{blockCommitted})$ message to $\mathcal{E}$, which means all honest validators commit on the same block for the same epoch (\textit{safety}).
Furthermore, the $(\texttt{blockCommitted})$ message is sent within $O(\tau+\Delta)$ time in optimistic cases and within $O(f(\tau+\Delta))$ time in pessimistic cases.
The \textit{weak consistency} is ensured by the \textbf{Challenge} functionality sending the $(\texttt{checkpointRevoked})$ message to $\mathcal{E}$, where the invalid unsettled checkpoint is revoked within $\Delta$ time.

\textit{Guaranteed fair withdrawal/exit.}
The \textit{liveness} is achieved by the \textbf{Transaction} functionality producing a settled checkpoint within $f(\tau+\Delta)$ time.
The two interfaces, claim and redemption, guarantee that during the period between two messages $(\texttt{accountFozen})$ and $(\texttt{fundsRedeemed})$ are sent to $\mathcal{E}$, the requester cannot spend the money on $\mathbb{SC}$, which prevents double spending.
The withdraw/exit procedure depends on settled checkpoints, thus the time complexity is also $O(f(\tau+\Delta))$.

\subsection{The \textsf{Sidechain} Protocol}

The \textsf{Sidechain} protocol is constructed exactly following the algorithms presented in Sec.~\ref{sec:sidechain}.
The protocol mainly consists of functionalities of the smart contract $\mathbb{C}$ and validators $\mathcal{V}$ (functionalities of clients $\mathcal{C}$ are contained).

We now present a brief summary of the \textsf{Sidechain} functionalities by describing functionalities of $\mathbb{C}$.
To transfer money from $\mathbb{MC}$ to $\mathbb{SC}$, the clients send request to the \textbf{deposit} functionality of $\mathbb{C}$.
Validators submit checkpoints by invoking the \textbf{submit} functionality, which only receives the checkpoint from the current leader.
The \textbf{withdraw/exit} functionality has two interfaces, claim and redemption, to ensure the fairness of withdrawal/exit.
Validators can challenge the unsettled checkpoint via the $(\texttt{challengeCP})$ interface of the \textbf{challenge} functionality, and trigger the view-change via the $(\texttt{challengeTimeout})$ interface.
When the validity of the smart contract expires, validators can terminate the \textsf{Sidechain} by invoking the \textbf{terminate} functionality.

\begin{figure*}[h!]

\begin{myFrame}{Functionalities of $\mathbb{C}$}

Assume $\mathbb{C}$ stores the current epoch number $epoch:=k$, current leader $L_k$, the last leader $L_{k-1}$, the last received checkpoint $cp_{k-1}$, and view-change timer $\Lambda$.

\begin{center}
	(I) \underline{\textbf{Deposit}}
\end{center}

A. Upon receiving $(\texttt{depositRequest}, \text{id}_{C_i}, x)$ from a client $C_i$:

\begin{enumerate}
    
    \item Request to $\mathcal{L}$ to transfer $x$ from $C_i$ (in $\mathbb{MC}$), by sending the message $(\texttt{depositRequested},\text{id}_{C_i},x)$ to $\mathcal{L}$.
    \item If $(\texttt{cannotDeposit},\text{id}_{C_i},x)$ is received, send $(\texttt{depositNotReceived},x)$ to $C_i$ and terminate the operation.
    \item If $(\texttt{depositConfirmed},\text{id}_{C_i},x)$ is received, send $(\texttt{depositReceived},x)$ to $C_i$ and to all the validators, for them to update the balance, and terminate the operation.
    
\end{enumerate}

\begin{center}
	(II) \underline{\textbf{Submit}}
\end{center}

A. Upon receiving $(\texttt{submit}, cp_k)$ from any of the validators, $V_i$, if $V_i=V_l$: 
\begin{enumerate}
    \item Reset the timer $\Lambda$, randomly select a new leader $V_j$, increase the epoch number $epoch:=k+1$ and all the epoch counters $\Upsilon:=\Upsilon+1$ for any clients, and set $cp_{k-1}.status:=settled$ and $cp_k.status:=unsettled$.
    
\end{enumerate}

\begin{center}
	(III) \underline{\textbf{Withdraw/Exit}}
\end{center}

A. Upon receiving $(\texttt{claim})$ from a client $C_i$:

\begin{enumerate}
    
    \item Set a epoch counter $\Upsilon_{C_i}:=0$ for $C_i$.
    
\end{enumerate}

B. Upon receiving $(\texttt{redeem}, x, PoP)$ from a client $C_i$, if $\Upsilon_{C_i}=3$ and $cp_{k-1}.status=settled$:
\begin{enumerate}
    \item Verify the provided $PoP$ (Proof of Possessions).
    If valid, transfer $x$ coins to the ledger $\mathcal{L}$ within $\Delta$ time.
\end{enumerate}

\begin{center}
	(IV) \underline{\textbf{Challenge}}
\end{center}

A. Upon receiving $(\texttt{challengeCP})$ from $V_i$:

\begin{enumerate}
    
    \item Run the verification of $cp_{k-1}$.
    If invalid, update the validators set $\mathcal{V}:=\mathcal{V} \setminus \{L_{k-1}\}$, set $cp_{k-1}:=\bot$.

\end{enumerate}

B. Upon receiving $(\texttt{challengeTimeout})$ from $V_i$ if the timer $\Lambda$ has expired:
\begin{enumerate}
    
    \item Reset the timer $\Lambda$, randomly select a new leader $V_j$, increase the epoch number $epoch:=k+1$ and all the epoch counters $\Upsilon:=\Upsilon+1$ for any clients, and set $cp_{k-1}.status:=settled$ and $cp_k:=\bot$.

\end{enumerate}

\begin{center}
	(V) \underline{\textbf{Terminate}}
\end{center}

A. Upon receiving $(\texttt{noClients})$ from a validator $V_k$:
\begin{enumerate}

    \item If the remaining balance $b=0$, send $(\texttt{executionEnd},1)$ to $\mathcal{E}$ and halt.
    \item If the remaining balance $b\neq 0$, send $(\texttt{executionEnd},0)$ to $\mathcal{E}$ and halt.
    
\end{enumerate}

\end{myFrame}
\caption{The functionalities of the smart contract $\mathbb{C}$}. \label{fig:smart-contract-functionality}
\end{figure*} 

We refer the readers to Appx.~\ref{appx:UCSmodel} for a detailed description of the UCS model of the protocol and Appx.~\ref{appx:security_proof} for the security proof.

\section{Implementation and Evaluation} \label{sec:implementation}

In order to evaluate the feasibility and performance of the protocol, we construct a simple proof of concept implementation of Cumulus.
The smart contracts are written in Solidity (around 800 LoC) and deployed in Ethereum Kovan testnet.
We also implemented the code of the validator side using Golang (around $1,900$ LoC) and the client side using Python (around $800$ LoC).

\subsection{Basic Performance}

We first evaluate the basic performance of Smart-BFT with $20$ replicas tolerating $f=1$ fault with a synchrony bound of $\delta=50$ ms.
We also implement PBFT using the same codebase of our Smart-BFT as the baseline.
Here we would like to point out that the goal of Cumulus is for off-chain scaling other than high performance and using PBFT for comparison is not entirely fair since PBFT does not assume synchrony.
Nevertheless, this kind of evaluation still helps understand to what extent can Smart-BFT achieve in terms of performance.
The two protocols were implemented in Golang and tested on the Amazon EC2 instances.
Each instance contains two virtual CPU cores and $8$ GB of memory.
Each transaction is a simple plain text transaction with zero payload and the block size is fixed to $10,000$ transactions.
We mentioned in Sec. ~\ref{sec:batching} that in order to push the performance of Smart-BFT to optimal, we can use some batching techniques.
Therefore, we consider different batch sizes, $60$, $100$, and $120$.

\begin{figure}[h!]
\begin{center}
  \includegraphics[width=0.9\linewidth]{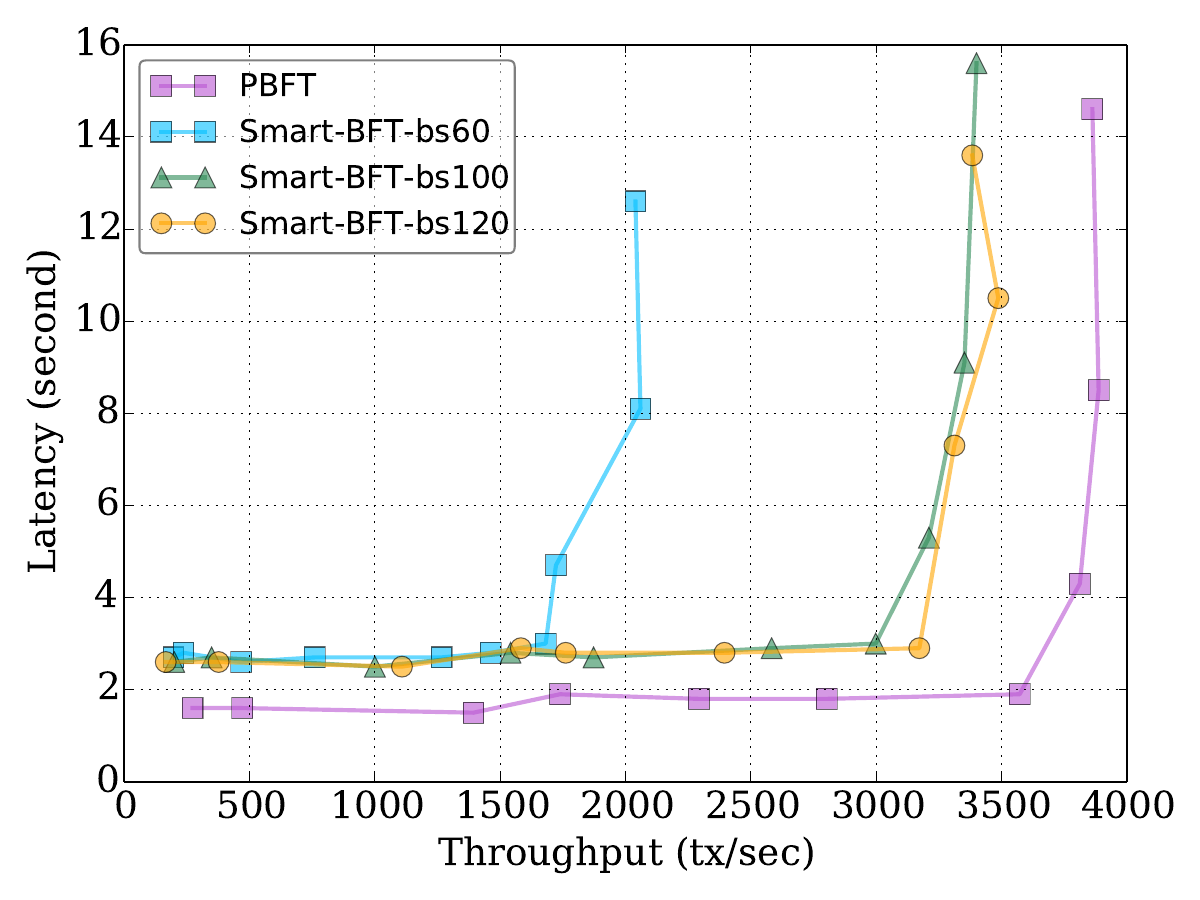}
  \caption{The performance of Smart-BFT with different batch sizes and with PBFT as the baseline.}
  \label{fig:throughput}
\end{center}
\end{figure}

Our results showed in Fig.~\ref{fig:throughput} indicate that the throughput of Smart-BFT is saturated at around $3,300$ tx/sec with a batch size of $400$ and there is no further throughput gain when batch size increases.
This means after the sidechain processes around $100$ blocks, a transaction containing a checkpoint can be finalized with high probability on the mainchain  (i.e., around $300$ seconds in the Ethereum according to public data available on https://etherscan.io/).
We can also observe that although PBFT slightly outperforms Smart-BFT due to different communication assumptions\footnote{Not that partially synchronous protocols usually perform better in a benign network since each node does not need to wait for $\delta$ in each round.}, the results still show that Smart-BFT is comparable with mainstream BFT-based protocols.
If all of the Ethereum transactions are taken up by sidechain checkpoints (imagine thousands of sidechains), even if only $10\%$, $\times 10,000$ scaling can be achieved.
We will compare Smart-BFT with more synchronous protocols in future work.

\subsection{Monetary Cost}

An important criterion to evaluate smart contracts running over the Ethereum is the cost of \textit{gas}, which is the fundamental unit of computation in the Ethereum.
The fees of any given fragment of programmable computation are universally agreed in the unit of gas.
Thus we can use the amount of gas to fairly measure the cost of each transaction.
Note that we only give the run-time cost of our protocol without considering the deployment cost because it is easy to optimize gas costs by extracting all  functionality of the smart contract into an external library.

\subsubsection{Constant Gas Cost}

There are two types of transactions, \textit{deposit} and \textit{checkpoint}, having constant gas cost, which means no matter how the number of validators and clients increases, the gas cost of these transactions remains unchanged because they send the constant size of data.
The exchange rate of gas to Ether is determined by GasPrice, which is freely chosen by the sender of the transaction.
Higher GasPrice means more incentive for miners to mine that transaction, which leads to shorter latency.
In our experiment, we choose an average rate, GasPrice$=4$ gwei (1 gwei $=10^{-9}$ Ether), under which the mean time to confirm a transaction is approximately $5$ minutes.
We use the exchange rate of $170$ between Ether and US dollar\footnote{According to the exchange rate as of 31 October, 2019}.

\begin{table}[!h]
\centering
\caption{Constant costs of deposit and checkpoint.}
 \begin{tabular}{|c|c|c|c|c|} 
 \hline
 Type & Bytes & Gas & ETH (gwei) & USD \\ [0.5ex]
 \hline\hline
 deposit & 4 & 21,912 & 87,600 & 0.015  \\ [1ex]
 \hline
 checkpoint & 98 & 307,206 & 1,089,300 & 0.207 \\ [1ex]
 \hline
\end{tabular}
\label{table:gascost}
\end{table}

Constant costs of deposit and checkpoint are list in Table~\ref{table:gascost}.
We observe that it only takes 0.015 USD to issue a deposit and the gas cost 21,912 is slightly more than that of a basic transaction (the gas cost of a basic value-transfer transaction is fixed to 21,000).
A checkpoint consists of the hash of the block header (32 bytes), two signatures (one of them is an aggregate signature, 128 bytes) and an index number (1 byte).
And it takes 307,206 gas (approx. 0.207 USD) for a leader to submit a checkpoint.
The duration of an epoch can be adjusted according to the actual use case.

\subsubsection{Withdrawal/exit cost}

\begin{figure}[h!]
\begin{center}
  \includegraphics[width=0.9\linewidth]{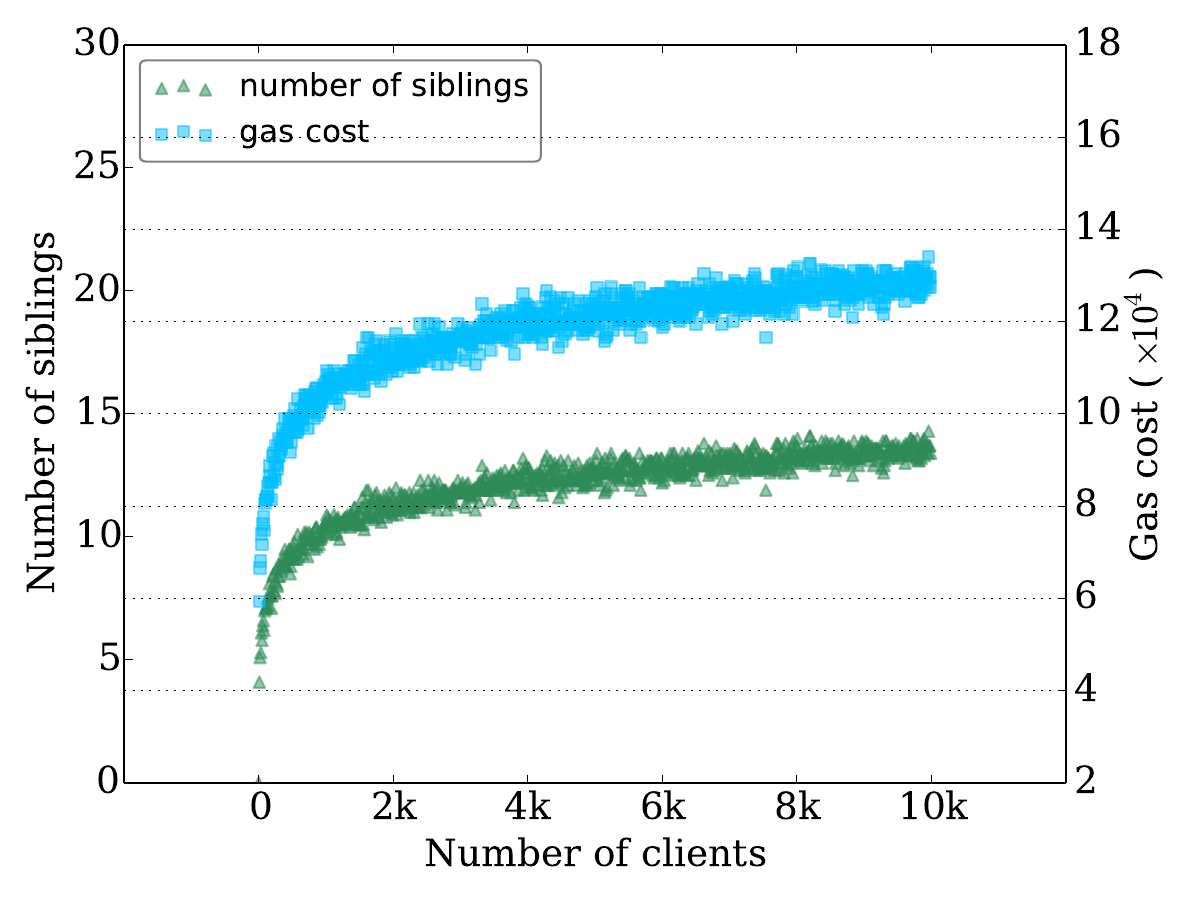}
  \caption{Gas cost of a \textit{redeem} transaction as the number of clients increases.}
  \label{fig:withdraw}
\end{center}
\end{figure}

Each withdrawal/exit request contains two on-chain transactions for \textit{claim} and \textit{redeem}.
The \textit{claim} transaction costs constantly, which is the same as the \textit{deposit} transaction, 21,912 gas (approx. 0.015 USD).
While as the number of clients scales up, the cost of a \textit{redeem} transaction increases because each client needs to submit a longer merkle path as the proof of possessions (Merkle proof) of the withdrawal/exit.
Each \textit{redeem} consists of a key (the address of the requester, 20 bytes), a value (the balance, 4 bytes), a root hash (the hash of the merkle trie, 32 bytes), a branch mask (a ``roadmap'' to the root node, 1 byte), and an array of siblings (adjacent nodes along the merkle path, 32 bytes).
The number of siblings grows with $m$ clients in $O(\log m)$, which is the major cost when the number of the clients becomes large.
We measured the \textit{redeem} cost as the number of clients grows in incremental steps (cf. Fig.~\ref{fig:withdraw}).
During each step, we randomly instructed one client to issue a \textit{redeem} transaction to $\mathbb{C}$, recorded the number of siblings included in the transactions and the gas cost.

We can observe that the gas cost grows in the same pattern as the number of siblings grows.
During the first 2,000 steps, the gas cost surges to around 80,000 (approx. 17 siblings) and then, for the rest of the steps, the speed of growth slows down.
When the number of clients grows to 10,000, the gas cost only increases to around 90,000 (approx. 20 siblings).
Using the same GasPrice and the exchange rate, we can calculate that it only costs about 0.078 USD for a user to withdraw/exit money from a sidechain network with 10,000 clients.
Based on the above observation, we can easily estimate the cost of withdrawal/exit when the number of clients scales up to one million.
For example, if one million clients joined the sidechain, a client would include 20 siblings in a withdrawal/exit transaction, which would cost 140,000 gas approximately (approx. 0.095 USD).

\subsubsection{Aggregate verification cost}

Our protocol applies a ``lazy-challenge'' scheme where $\mathbb{C}$ will not run the verification of the pending checkpoint until some party submits a ``challenge'' request.
There are two types of challenges: (i) a challenge denoted by \textit{challenge-ncp} complaining the leader for not committing the checkpoint after a timeout, (ii) and another challenge denoted by \textit{challenge-cp} complaining the unsettled checkpoint (faulty aggregate signature).
The former requires parameters to include an aggregate signature assembled by the challenger and an index number, while the latter requires no parameters.
Both types of challenges will trigger $\mathbb{C}$ running the verification of the aggregate signature.
We implemented BGLS aggregate signature \cite{boneh} in Solidity based on an existing codebase\footnote{https://github.com/Project-Arda/bgls-on-evm}.

\begin{figure}[h!]
\begin{center}
  \includegraphics[width=0.9\linewidth]{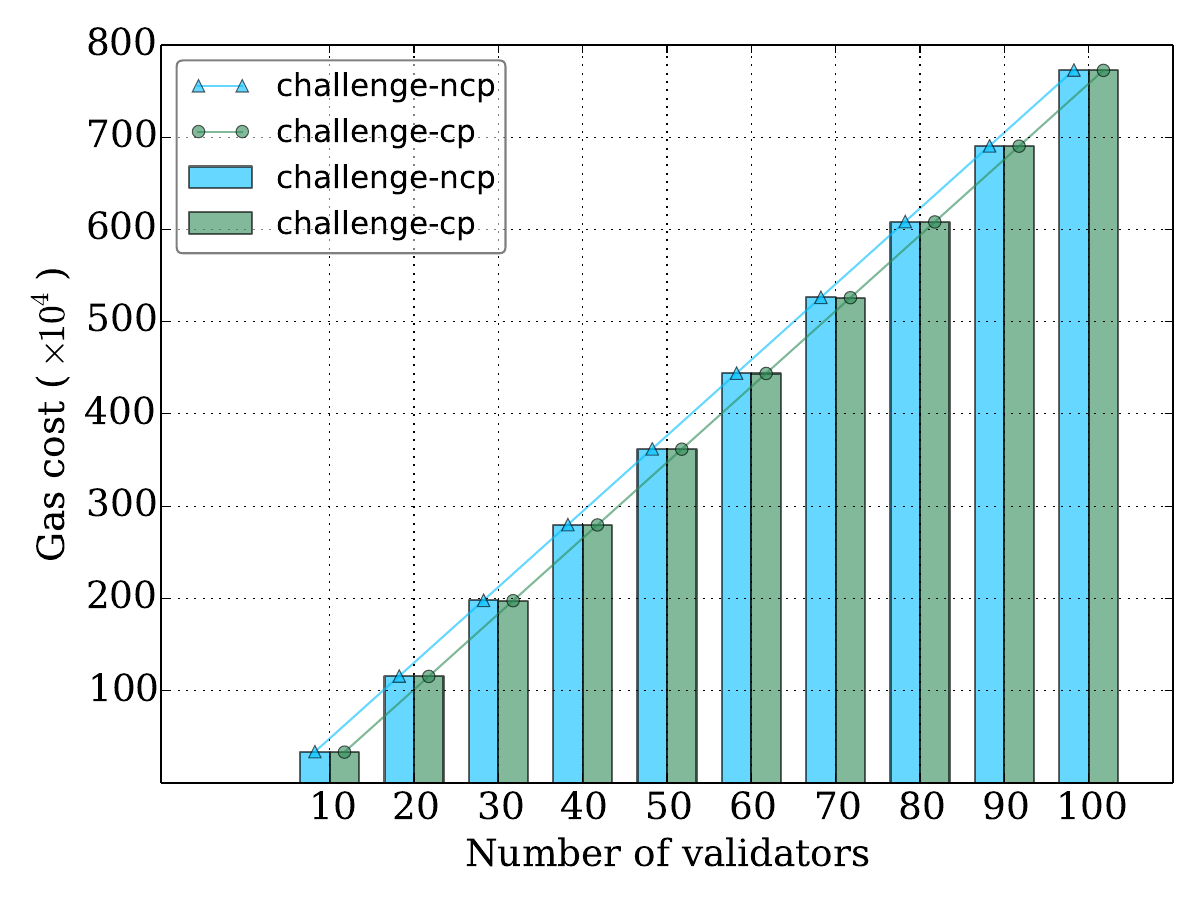}
  \caption{Gas cost as the number of validators increases.}
  \label{fig:verification}
\end{center}
\end{figure}

Fig.~\ref{fig:verification} showed the gas cost of the aggregate verification as the number of validators increases.
We can see that the gas cost (triggered by \textit{challenge-cp}) grows linearly with the validator number from $334,659$ with $10$ validators to $7,723,091$, approx. $0.208$ USD to $4.8$ USD (gas cost triggered by \textit{challenge-ncp} from $338,953$ with $10$ validators to $7,727,248$, approx. $0.0.211$ USD to $4.802$ USD because of additional parameters).
This is because when the number of validators grows, the complexity of the verification increases linearly.
Note that the validator number is limited even though we ignore the monetary cost because there is a \textit{gas limit} which defines all the transactions inside a block allowed to consume in Ethereum.
However, the \textit{gas limit} can be adjusted by the miner due to the congestion, which is about $8,000,000$ gas on average.

\section{Related Work} \label{sec:related-work}

The idea of off-chain scaling originated from \textit{Lightening Network} \cite{poon2016bitcoin}, which has emerged as the most promising PCN practice for Bitcoin, followed by \textit{Sprites} \cite{miller2017sprites} and \textit{Raiden} \cite{raiden} for Ethereum.
There exists extensive literature proposing constructions to improve PCN.
Revive \cite{revive2017}, Spider \cite{Sivaraman_2018} , and Flash \cite{wang2019flash} propose dynamic routing algorithms to maximize the throughput and success volume of PCN, while Perun \cite{perun2019} introduces \textit{virtual payment channels} to avoid the involvement of the intermediary for each payment, thereby, significantly reducing communication complexity.

In contrast to channels between two parties, sidechain-based approaches allow parties in a sidechain network to freely perform transactions without a prebuilt channel.
The first sidechain-based scheme that enables off-chain transactions is Plasma \cite{poon2017plasma}, which allows arbitrary consensus algorithms in the sidechain.
The consensus mechanism enforces the rules encoded in the smart contract, through which disputes can be resolved fairly.
Therefore, as long as the mainchain remains secure, the safety is ensured in the sidechain even if the consensus mechanism is compromised.
Two main challenges of current plasma constructions \cite{plasmamvp}\cite{plasmacash} are the size of proofs and cumbersome withdrawal/exit procedure.

A very interesting construction similar to ours is called NOCUST \cite{rami2018nocust}, which is an account-based  \textit{commit-chain}, where a single operator maintains users' account and commits checkpoints at regular intervals.
NOCUST is a challenge-response protocol, where users can issue different types of challenges directly to the smart contract to mediate.
As such, it requires users to be online at least once within a block time, or a malicious operator might manipulate users' balances.
NOCUST also utilizes zkSNARK \cite{zkSNARK} in its smart contract which enables efficient verification of the complete correctness of checkpoints.
A major difference between NOCUST and Cumulus is that our protocol features a BFT-based consensus that ensures safety and liveness without online requirements under appropriate security assumption.
We argue that by introducing a consensus mechanism over even a small-sized committee in the sidechain can mitigate single-point-failure and improve user experience.

Other related works include proof-of-stake (PoS) sidechains \cite{pos-sidechain} which can be adapted for other PoS blockchain systems.
Platypus is an off-chain protocol for blockchains, which is a \textit{childchain} that requires neither synchronous assumptions nor a trusted execution environment.
However, these two sidechain protocols are mainly tailor-made to their own designed blockchains, instead of public blockchains such as Bitcoin and Ethereum.

It should be noted that there have been recent work on improving the performances of partially synchronous BFT protocols \cite{SBFTProtocol,WindowBasedBFT,ConsistentBFT,hotstuff,ExperimentalEvaluationOfBFT,Proteus,DBLP:conf/podc/YinMRGA19} but these protocols can only tolerate less than one third $f \leq \frac{n-1}{3}$ Byzantine nodes in the network. Therefore, we used Synchronous Byzantine Fault Tolerant protocol which not only matches our use case but also can tolerate less than half $f \leq \frac{n-1}{2}$ Byzantine nodes.

Recently, UCS framework has been widely used to prove the security of off-chain solutions\cite{miller2017sprites,perun2019,general,fair-swap,multyparty,malavolta2019anonymous}.
Given the serious monetary loss when off-chain protocols are massively deployed, we believe that the security of each off-chain protocol should be formally proved.

\section{Conclusions} \label{sec:conclusion}

In this paper, we introduced Cumulus, a sidechain-based off-chain protocol for public blockchains to achieve scale-out throughput without sacrificing security.
We developed a new consensus algorithm, called Smart-BFT, customized for sidechains which ensures liveness and safety with a maximum of $f=\lfloor\frac{n-1}{2}\rfloor$ validators being Byzantine.
Users in Cumulus can enjoy instant finality, near-zero transaction fee, fair withdrawal/exit without online requirements.
We defined the security of our protocol in the UC framework and provided formal proofs.
The performance evaluation showed that Cumulus has the potential to scale the underlying blockchain to the order of $10,000 tx/s$ for throughput and several seconds for latency.
Cumulus is also practical and scalable in terms of monetary cost: around $0.207$ USD for each checkpoint and $0.095$ USD for each withdrawal in the order of million users.
For future work, we plan to extend our sidechain protocol into a partially synchronous model and do more performance comparison with more mainstream BFT-based protocols.

\bibliographystyle{IEEEtranS}

\bibliography{usenix2019_v3.1}

\appendix

\section{UCS Framework Model} \label{appx:UCSmodel}

\begin{figure}
\begin{center}
  \includegraphics[width=0.7\linewidth]{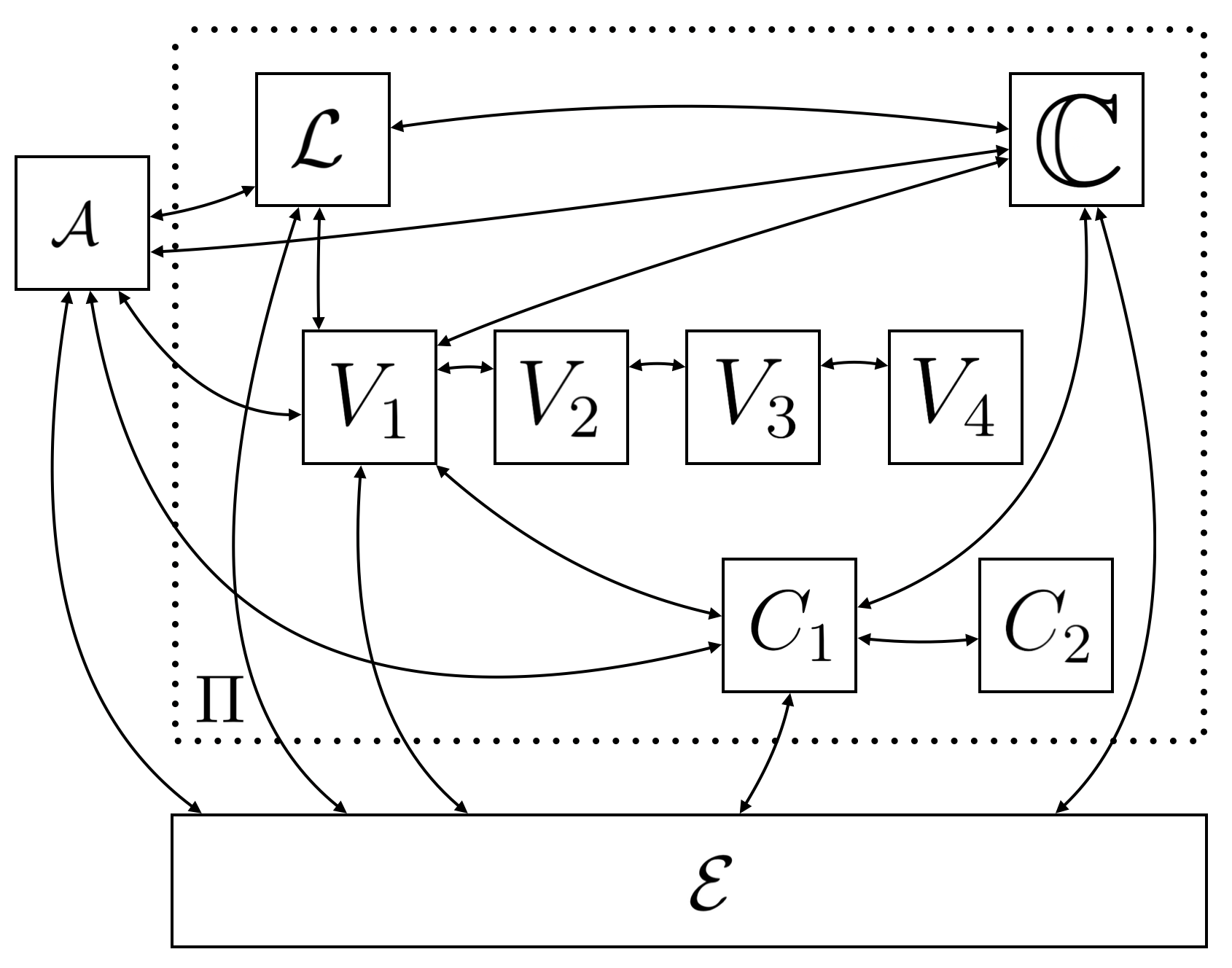}
  \caption{Protocol $\Pi$ model.}
  \label{fig:Protocol_Pi}
\end{center}
\end{figure}

The protocol $\Pi$ is built with $4$ computing entities, validators $\mathcal{V}$, clients $\mathcal{C}$, the mainchain (or global ledger) $\mathcal{L}$, and the smart contract $\mathbb{C}$ to perform different functions.
Each one of these entities is identified with a unique ID.
The basic construction of the protocol is depicted in Fig.~\ref{fig:Protocol_Pi}, where only all the communication paths for $V_1$ and $C_1$ are depicted for simplicity, showing $2$ clients and $4$ validators.

The following is the functional description of each of the elements listed above.
As we describe them, we present all the parameters that they require to work and the mathematical notation for them.

\subsection{Validators}

The validators are nodes that clients request transactions to.
They are responsible for collecting and storing the transactions that the clients send and compile the blocks that contain the accumulated transactions during the epochs, for other validators to collectively approve.

During the execution of the protocol, validators can take one of two different roles: leader or follower.
When a certain amount of transactions are accumulated, or a certain amount of time has passed (parameters that we can assume are fixed prior to the start of the protocol), one of the validators takes the responsibility of carrying out the process of compiling the transactions into a block.
This validator is the leader of the epoch.
The rest of the validators are followers, which are only responsible for receiving the block compiled by the leader, and approve it if the information contained corresponds to its internal registry, or disapprove if not. 

The validators' parameters are the following.
First, there are two read-only numbers, that are the maximum amount of transactions per epoch, $n_{\max}$, and the maximum time duration for an epoch $t_{\max}$. There is an index number $v_{index}$, which has the order number in which the particular validator is picked to be leader, and a number $v_n$ that gives how many validators are in the sidechain. There is also a binary flag $v_{status}$ which indicates the status of the validator, $1$ if it is the epoch leader or $0$ if is a follower. For synchronization purposes, all validators are aware of the current leader by $v_{current}$, this parameter is set to 1 at the beginning of the execution, increases whenever an epoch starts, and goes back to 1 if it goes greater than $n$. Whenever $v_{current}=v_{index}$, then $v_{status}=1$, otherwise $v_{status}=0$. The validators control the number of transactions of the current epoch with $n_{txn}$, and $t_{txn}$ is a timer with the duration of the current epoch; both numbers are set to 0 at the beginning of the epoch. The validators also hold $\mathcal{P}^* = \{\text{id}_{P_1}, \text{id}_{P_2}, \dots , \text{id}_{P_m}\}$, the set of IDs of the clients. Also, a function $B: \mathcal{P}^* \times \mathbb{N} \rightarrow \mathbb{N}$, that gives the balance of the corresponding client at an epoch $e$, i.e., $B(\text{id}_{C_i}, e) = b_{i,e}$, where $b_{i,e}$ is the balance of the client $C_i$ at epoch $e$. Among the validators, the balances function (which can be viewed as an array), must be equal. We denote individual transactions (plain text transactions) as a tuple, $(C_i,C_j,b)$, where $C_i$ is the sender, $C_j$ is the receiver and $b$ the amount. Actual transactions require a digital signature to ensure that they legitimately come from the client $C_i$, so the transactions that are sent to validators are also tuples, denoted by $tx_l = (C_i,C_j,b,s_{i,j,b})$, where $s_{i,j,b}$ is the digital signature. They also hold the transactions queue $Q_{txn}(e)$ (for the epoch $e$), that at the beginning of each epoch is an empty set. As transactions are issued by clients, they are appended to this queue, so it is of the form $Q_{txn}(e) = \{tx_{1}, tx_{2}, \ldots  , tx_{l}\}$. Lastly, the validators hold $b^s_i$ and $b^r_i$, that are  a pair as accumulators for the received and sent amounts for each client during an epoch. At the beginning of an epoch, both are equal to zero.

\subsection{Clients}

The clients are the nodes that join the protocol and make use of $\mathbb{SC}$ to make off-chain money transfers to other clients. When a new client wants to join $\mathbb{SC}$, it has to make an initial deposit to $\mathbb{C}$ and identify itself with an ID that is not already part of $\mathbb{SC}$. Also, clients that have a positive balance on $\mathbb{SC}$, can withdraw their money to have it returned to $\mathbb{MC}$. When a client that is already part of the network, wants to leave $\mathbb{SC}$, it has to request its exit, which automatically issues withdraw for the total of its remaining balance, then the ITM that represents the client goes to a halt state. 

The clients are responsible of keeping track of their own balances, and query from the smart contract the necessary elements (Merkel path) to provide proof of possession, whenever any of them request to withdraw from or exit $\mathbb{SC}$.  

For this scheme to work, each client requires to have a function $B_i: \mathbb{N} \rightarrow \mathbb{N}$ that gives its own balance for any epoch, i.e., $B_i(e) = b_{i,e}$, where $b_{i,e}$ is the balance at epoch $e$ for client $C_i$. We assume that if a client joined at an epoch $k$, (i.e., it did not join since the beginning of the execution), then $B_i(k')=0$ for all $k'<k$.

\subsection{Global Ledger}

The Global Ledger $\mathcal{L}$ is the computing entity that represents the functionality of $\mathbb{MC}$. It receives the block headers of each of the transaction batches, to serve as checkpoints for transactions that occur on $\mathbb{SC}$. For our purposes, we can assume that clients have infinite off-chain balance since the security of the mainchain is outside of the scope. 

\subsection{Smart Contract}

The Smart Contract $\mathbb{C}$ is the computing entity that controls the execution of the protocol by having a counter $e$ that holds the current epoch, this counter starts in 1 at the beginning of the execution. Also, it holds a function that controls the total money that is flowing in the protocol at any epoch, $b_t: \mathbb{N} \rightarrow \mathbb{N}$, so $b_t(e) = \sum_{i} b_{i,e} = \sum_{i} B(\text{id}_{C_i},e)$. It also holds the set of IDs for the validators $\mathcal{V}^* = \{\text{id}_{V_1}, \text{id}_{V_2}, \dots , \text{id}_{V_n}\}$. 

\begin{figure}
\begin{center}
  \includegraphics[width=0.7\linewidth]{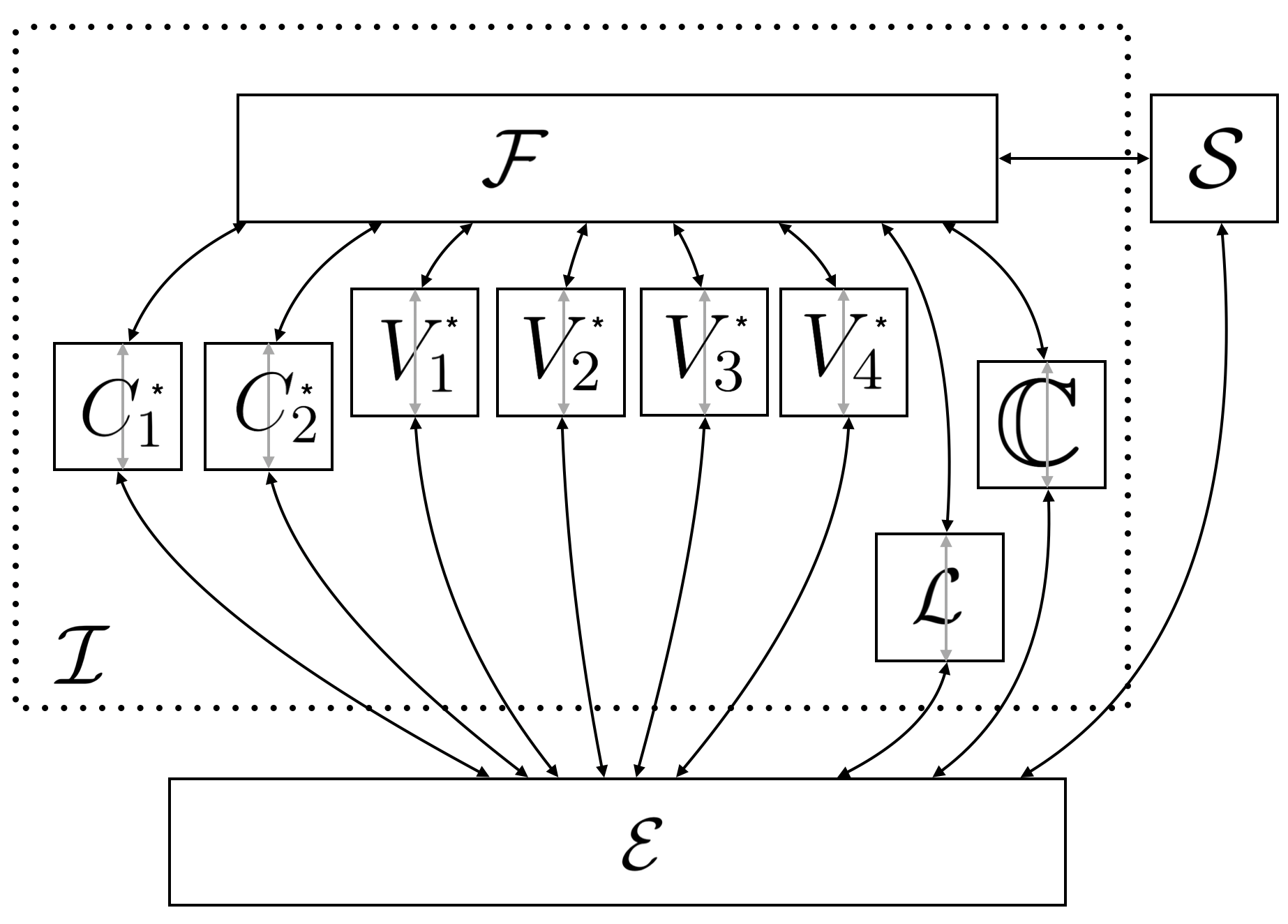}
  \caption{Ideal protocol $\mathcal{I}$ model.}
  \label{fig:Protocol_I}
\end{center}
\end{figure}

The security proof relies on making to any environment, the real protocol indistinguishable from the ideal functionality $\mathcal{F}$ designed to model how the protocol works and presented in Sec.~\ref{sec:ideal-function}. Fig.~\ref{fig:Protocol_I} depicts the ideal protocol. 


\section{Security Proof} \label{appx:security_proof}

For the security proof we will use as a basis the formal definitions of the protocol from Sec.~\ref{sec:security}, by showing that for any adversary $\mathcal{A}$ interacting with the real protocol (Fig.~\ref{fig:Protocol_Pi}), we can always find a simulator $\mathcal{S}$ such that the environment machine cannot distinguish if it has interacted with the real protocol $\Pi$ or the ideal protocol $\mathcal{I}$. The concept of distinction of one protocol from the other comes from a vanishing probability of getting different outputs on the environment machine, after running with $\Pi$ or $\mathcal{I}$.

A simplification regarding the full UCS model is that we do not make use of session IDs since we consider all interactions to outside entities through the environment $\mathcal{E}$, i.e., no machine from the protocol can communicate to outside machines. 

For the proof, we are going to use a constructive approach, this is, we are going to take the computing entities we described in Appx.~\ref{appx:UCSmodel}, and evaluate how the protocol works interacting with it (as depicted in Fig.~\ref{fig:Protocol_Pi})

The UCS model requires that at any point of the execution, only one machine is active, which means the protocol operates with a sequential execution and whenever a machine sends input to or requests output from another machine, it stops its execution and the other machine activates upon getting the input or upon being requested to compute some output.

To this end, we will consider that the flow of the execution of the protocol is controlled by the environment as follows. First, the execution starts when $\mathcal{E}$ requests a deposit for a new client, and there are no machines yet (no validators, clients, smart contract or global ledger); the input parameter for the execution is just the amount for this deposit. When this happens, $\mathcal{E}$ invokes the smart contract $\mathbb{C}$ machine. Then $\mathbb{C}$ proceeds to process the deposit, but as there are no global ledger or validators, $\mathcal{L}$ and the set of $n$ validators $\{V_1, \ldots, V_n\}$ are invoked at the moment they are required ($n$ being a fixed predefined parameter by $\mathcal{E}$). When the deposit is confirmed by $\mathcal{L}$,  and the first validator is requested to confirm the deposit to the first client $V_1$, it invokes the machine for it. Whenever a client or a validator is invoked, it generates its own pair of private and public keys.

As described before, new clients are invoked by validators but triggered by $\mathcal{E}$, by requesting deposits with IDs that are not in the client set. The environment, thus, is aware of all the clients that are in the sidechain and the order in which they joined. We consider then, that the environment sequentially, and in the same order, sends messages to the clients asking if they require to make any movement (deposit, withdraw, transaction or exit). With this, we keep the sequential nature of the protocol, while having the scheme working as described before.

We also consider that the protocol execution finishes when there is just one client left and it requests to exit. Lastly, the output of the environment machine is as follows: 1 if the last withdraw to exit is successful (this is, the last client left the sidechain by providing a correct proof of possession for the remaining balance on the smart contract), and 0 if this withdrawal is not successful, since under these circumstances, at some point something went wrong. The output of the environment machine is the value that is written to its output tape when it halts.

Recall the security assumption is that at most $\lfloor\frac{n-1}{2}\rfloor$ validators could be potentially dishonest. First, we consider the case when the protocol running is $\mathcal{I}$ against an adversary $\mathcal{A}$.

The ideal functionality $\mathcal{F}$ keeps $n$ internal registries, one for each validator. Considering the security assumption, only at most $\lfloor\frac{n-1}{2}\rfloor$ of these registries can be tampered by the action of the adversary $\mathcal{A}$, through backdoor messages to $\mathcal{F}$. Now, to avoid testing the protocol against all possible adversaries $\mathcal{A}$, it suffices to make it interact with the ``dummy adversary'' $\mathcal{D}$, since if a protocol $\mathcal{I}$ UC-emulates another protocol $\Pi$ with respect to the dummy adversary, then $\mathcal{I}$ UC-emulates $Pi$ against any adversary \cite{canetti2001universally}. The dummy adversary, basically allows $\mathcal{E}$ to take over the communication links between the machines, this is done by acting as a buffer between $\mathcal{E}$ and all clients in the protocol: if $\mathcal{D}$ receives a message from $\mathcal{E}$ to be delivered to any participant, it sends it as a backdoor message to the intended participant. And if $\mathcal{D}$ receives a backdoor message, it delivers it to  $\mathcal{E}$. This is, if $\mathcal{E}$ wants to attack any participant, it does so via $\mathcal{D}$, and any response that  $\mathcal{D}$ gets is forwarded to $\mathcal{E}$. Such attacks, in general, are messages to modify any of the information flowing in the protocol: transactions, deposits or withdraws amounts. 

The internal registries are, however, protected by the security features that the real protocol counterpart: the block is still composed using hashes, and transactions are still signed. A consequence of this, the probability of finishing the execution of the protocol with a zero output is null: 

Assuming that at most $\lfloor\frac{n-1}{2}\rfloor$ registries have been compromised by the adversary, the block proposal process could be carried out with compromised information. However, the verification process has to check among the $n$ registries, and after finding inconsistencies in more than half, the block would be discarded. This is also true if one or more of the registries have been modified maliciously for the benefit of one or more malicious clients. This feature prevents double spending, which in turn would cause the protocol to output 0 upon fishing its execution.

We now turn our attention to the protocol $\Pi$. Let's consider any adversary $\mathcal{A}$. We are going to build a simulator $\mathcal{S}$ in the following way. $\mathcal{S}$ will internally run an instance of $\mathcal{A}$ and another machine, $\mathcal{S_{\mathcal{D}}}$. This machine will serve as an interface from $\mathcal{A}$ to send and receive backdoor messages to and from any participant of $\Pi$. $\mathcal{A}$ is connected to $\mathcal{E}$ to receive or send messages, and to $\mathcal{S_{\mathcal{D}}}$. Essentially: (1) When $\mathcal{S}$ is sent a message by $\mathcal{E}$, it runs the internal instance of $\mathcal{A}$. If the output message is intended for one of the participants, it is delivered to $\mathcal{S_{\mathcal{D}}}$, and it sends it to that participant as a backdoor. If the output message is intended to $\mathcal{E}$, it is delivered to it. (2) When $\mathcal{S}$ is sent a backdoor message by any of the participants, it is delivered to the internal instance of $\mathcal{A}$, and the resulting output is handled in the same way. 

With this construction, we have replicated a subset of functions of the dummy adversary: while $\mathcal{D}$ gives full control to $\mathcal{E}$ to perform any kind of attacks, $\mathcal{S}$ encompass only some of the attacks that may be possible, limited by what $\mathcal{A}$ can do.

As in the case of the ideal protocol, the execution is protected by the security properties defined before, so the probability of getting a 0 as an execution output is also zero. With this, it becomes indistinguishable for $\mathcal{E}$ to detect if the protocol that runs, was $\Pi$ or $\mathcal{I}$.

\end{document}